\documentclass[useAMS,usenatbib]{mn2e}

\usepackage{graphicx}
\usepackage{rotating}
\usepackage{longtable}
\usepackage{amsmath}
\usepackage{amstext}
\usepackage{amssymb}
\def   \aj {{\rm {AJ}}}
\def   \araa {{\rm {ARA\&A}}}
\def   \apj {{\rm {ApJ}}}

\def   \apjs {{\rm {ApJS}}}

\def   \aap {{\rm {A\&A}}}

\def   \mnras {{\rm {MNRAS}}}

\def   \apjl {{\rm {ApJL}}}

\def   \caa  {{\rm {CA\&A}}}

\title[Massive star forming cores]{A distance limited sample of massive star forming cores from the RMS\thanks{http://rms.leeds.ac.uk/cgi-bin/public/RMS\_DATABASE.cgi} Survey}
\author[L. T. Maud et al.]
{L. T. Maud$^{1,2}$\thanks{E-mail:maud@strw.leidenuniv.nl (LTM)},
  S. L. Lumsden$^{1}$, T. J. T. Moore$^{3}$, J. C. Mottram$^{2}$,
\newauthor J. S. Urquhart$^{4}$ and A. Cicchini$^{1,5}$\\
$^{1}$School of Physics and Astronomy, University of Leeds, Leeds, LS2 9JT, UK\\
$^{2}$Leiden Observatory, Leiden University, PO Box 9513, 2300 RA Leiden, The Netherlands\\
$^{3}$Astrophysics Research Institute, Liverpool John Moores University, 146 Brownlow Hill, Liverpool, L5 3RF, UK\\
$^{4}$Max-Planck-Institute f\"{u}r Radioastronomie, Auf dem H\"{u}gel 69, D-53121 Bonn, Germany\\
$^{5}$School of Physics and Astronomy, University of Nottingham, Nottingham, NG7 2RD, UK\\
}
\begin{document}

\date{Accepted 2015 June 12. Received 2015 June 12; in original form 2015 February 23}

\pagerange{\pageref{firstpage}--\pageref{lastpage}} \pubyear{2015}

\maketitle

\label{firstpage}

\begin{abstract}

 We analyse C$^{18}$O ($J=3-$2) data from a sample of 99 infrared-bright massive
 young stellar objects (MYSOs) and compact H{\sc ii} regions that were
 identified as potential molecular-outflow sources in the Red MSX source (RMS)
 survey. We extract a distance limited (D\,$<$\,6\,kpc) sample shown to be
 representative of star formation covering the transition between the source
 types. At the spatial resolution probed, Larson-like relationships are found
 for these cores, though the alternative explanation, that Larson's relations
 arise where surface-density-limited samples are considered, is also consistent
 with our data.  There are no significant differences found between source
 properties for the MYSOs and H{\sc ii} regions, suggesting that the core
 properties are established prior to the formation of massive stars, which
 subsequently have little impact at the later evolutionary stages
 investigated. There is a strong correlation between dust-continuum and
 C$^{18}$O-gas masses, supporting the interpretation that both trace the same
 material in these IR-bright sources. A clear linear relationship is seen
 between the independently established core masses and luminosities. The position 
 of MYSOs and compact H{\sc ii} regions in the mass-luminosity plane is 
 consistent with the luminosity expected a cluster of protostars 
 when using a $\sim$40 percent star-formation efficiency and indicates that they are at a
 similar evolutionary stage, near the end of the accretion phase.

\end{abstract}


\section{Introduction}
\label{seccoreINTRO}

Massive stars ($>$ 8\,M$_{\odot}$) are responsible for some of the most
energetic phenomena in the Galaxy. They deposit large amounts of radiation,
kinetic energy and enriched material into the interstellar medium (ISM)
throughout their formation, main-sequence lifetimes and when they explode as
supernovae. Massive young stellar objects (MYSOs) are the precursors to massive
stars and are luminous ($\gtrsim$ 10$^3$\,L$_{\odot}$), mid-infrared point
sources which have not yet begun to ionise their surroundings
\citep{Davies2010}. The details of their early formation stages are difficult
to probe observationally, due to their rare, clustered and embedded nature
\citep{Cesaroni2007}. As a result, the processes of massive star formation are
still relatively uncertain when compared to the well studied low-mass
star-formation paradigm \citep{Shu1987}.

Theoretical modelling suggests that, during their formation, MYSOs with high
accretion rates (10$^{-4}$ M$_{\odot}$\,yr$^{-1}$) swell due to the mass
influx. Consequently, they are deficient in ultraviolet (UV) photons until they
begin contracting to a near-Main-Sequence configuration
\citep{Hosokawa2009,Hosokawa2010}; thus despite being very luminous, they are
not initially ionising their surroundings. As the MYSOs contract towards the
main sequence, they will then start to ionise the surrounding environment and
form expanding H{\sc ii} regions \citep{Hoare2007}. When the central stars
reach the main sequence they generate copious amounts of UV photons, and so
rapidly disrupt and destroy the natal cloud. Identifying MYSOs and
very young H{\sc ii} regions \citep[very compact radio
  emitters,][]{Lumsden2013} provides a sample of sources in which the natal
environment will be less disrupted and therefore closer to their initial
conditions, while simultaneously facilitating the investigation of feedback
from the massive protostars known to be forming.  The Red MSX Source (RMS)
survey \citep{Lumsden2013} is an ideal sample for this since it spans a wide
range of luminosity and evolutionary stage.  It also allows us to study how
molecular gas properties vary as a function of both time and source luminosity.

This paper deals with the core properties of a sample of primarily northern RMS sources.
These observations constitute the only RMS dataset so far in which the molecular-gas emission
has been mapped, as opposed to obtaining single-pointing observations \citep[cf.
][]{Urquhart2011b}. The
primary goal of the observations was to study molecular outflows using single-dish 
observations of $^{12}$CO, $^{13}$CO and C$^{18}$O ($J=3-$2)
(discussed in a companion paper: Maud et al. 2015 submitted to MNRAS ). The aim of the
C$^{18}$O observations was to study the kinematical behaviour of the gas in the
molecular core around the RMS sources.  This allows us to determine outflow
properties more reliably, but also permits us to study the cores separately.
Single-dish C$^{18}$O (3$-$2) is generally optically thin \citep{Zhang2009} and
excited in the denser regions when compared with lower-excitation transitions of CO
(e.g. $J=1-$0). The $J=3-$2 emission is much less confused with
line-of-sight emission than lower-excitation transitions, and a critical
density for the transition of $>10^4$cm$^{-3}$ (with most emission expected to
come from significantly higher densities - cf. \citealt{curtis2011}), makes it an ideal tracer of
the core structures in star-forming regions.

Section \ref{obs_core} describes the source sample and observations undertaken, with Section
\ref{meth_cores} detailing the method used to calculate all source masses and
their radii. Section \ref{samp} presents the basic results and the comparisons with previous data
from the literature and the RMS survey itself. 
Section \ref{analysis} discusses various relationships with source properties and in
Section \ref{comp_lum_CO} we compare our results with a simple model to examine the star
formation efficiency and protostellar evolution. A summary is given in Section \ref{conc}.

\section{Sample and Observations}
\label{obs_core}
\subsection{Sample Selection}

The sources were chosen from all MYSOs and H{\sc ii} regions in the RMS survey
that are located within a distance of $\sim$6\,kpc, have luminosities $\gtrsim$
3000\,L$_{\odot}$, and are observable with the JCMT (declinations
$-$25$^{\circ}$ to $+$65$^{\circ}$), with some additional right ascension
constraints set by the observing dates.  In addition, for the H{\sc ii}
regions, only those sources which appear compact in higher resolution
mid-infrared images were selected.  Finally although all of the sources with
$L>10000$\,L$_{\odot}$ were observed, only a random sample of the less luminous
ones were included (see below). The original selection was made using the
pre-2008 RMS catalogue and resulted in 99 target sources representative of the
$\sim$200 MYSOs and compact H{\sc ii} regions satisfying these criteria in the
catalogue at that time. Since 2008, the RMS catalogue has evolved significantly
as a result of subsequent observations.  The updates to the RMS database have
resulted in 10 of the initial sources now being assigned a kinematic distance
beyond 6\,kpc, where luminosities are only complete to $\sim$10$^4$ L$_{\odot}$
\citep{Mottram2011b}; these are removed from the statistical sample (although
masses are still calculated where possible). The luminosities for all sources
have been recalculated using the most up-to-date source distances
\citep{Urquhart2012,Urquhart2014} and multi-wavelength spectral energy
distribution (SED) fits from \citet{Mottram2011a}.  As luminosities have
changed the sample now extends to $\sim$1000\,L$_{\odot}$, but is not complete
below the original limits.  Finally, there are now a total of $\sim$311 sources in the RMS catalogue
that satisfy the original distance-limited criteria described above.
Of these, we would expect about 50 H{\sc ii} regions to be too extended (
  based upon the RMS classifications), so that
a complete distance-limited sample from the final catalogue would number
approximately 260 sources.

\begin{figure*}
\begin{center}
\includegraphics[width=0.95\textwidth]{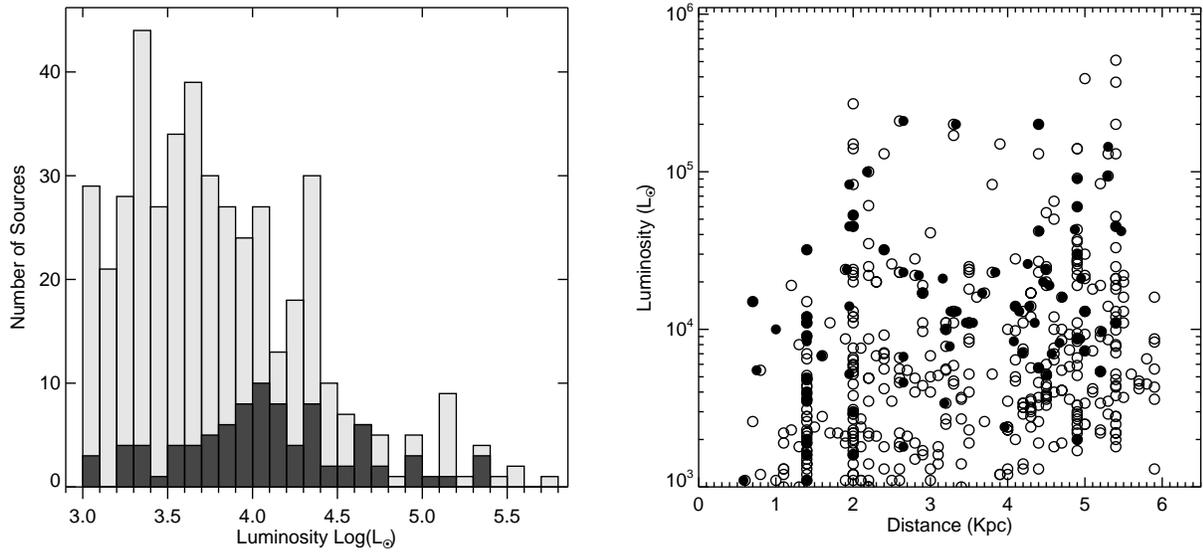}
\caption{Left: Luminosity distribution of RMS sources now conforming to the original sample criteria (light grey bars) over-plotted with the luminosity distribution of the distance-limited set of observed sources (dark grey bars). The range of luminosities is well covered.  Right: Comparison of the RMS distances and luminosities of the 89  sources in the distance-limited sample (filled circles) with the MYSOs and H{\sc ii} regions now conforming to the original sample criteria (open circles) (see text for details). The sample covers the range of luminosity in the present population of $\sim$450 RMS sources that 
now match the original criteria. Note that sources at the same distance form part of known molecular complexes. Uncertainties are not indicated although are $\sim$\,30\,percent for luminosity \citep{Mottram2011a} and an illustrative value of 50\,percent can be adopted for the distance.}
\label{fig:dist_lum} 
\end{center}
\end{figure*}

The left panel of Figure \ref{fig:dist_lum} shows the luminosity distributions
of the remaining 89 sources (dark grey) and of $\sim$450 objects in the RMS
database (light grey) now matching the original criteria (but where
L$>$1000\,L$_{\odot}$ and source types are MYSO and both compact and extended H{\sc ii} regions).
The right panel shows the luminosities as a function of
distance in the two sets of sources (filled and open circles,
respectively). Both plots indicate that the range of luminosities and range of
distances probed by the original observed sample is representative of all
sources in the RMS database that satisfy the selection criteria.  It is clear
that the observed sample is incomplete in terms of relative numbers of sources
at luminosities below $10^4$\,L$_{\odot}$ (above which the RMS survey is now
complete).  This was always intended however, since our initial sampling aimed
to have approximately equal numbers per logarithmic luminosity bin.  
Since all the analyses in this paper are essentially of ratios of observed quantities 
(e.g. Section \ref{comp_lum_CO}, where luminosity is concerned), this 
numerical incompleteness is not a significant problem.

\subsection{Observations}
All 99 sources were observed with the James Clerk Maxwell Telescope (JCMT) as
part of projects M07AU08, M07BU16, M08AU19 and M08BU18 during 2007 and
2008. The 15-m dish yields a full-width half-maximum (FWHM) beam size of
15.3\,arcsec at $\sim$329\,GHz for the C$^{18}$O (3-2) line. Throughout the
observations, the typical median system temperatures were $T_{sys}$
$\sim$350$-$550\,K. In a few cases, system temperatures reached as high as
900\,K, reflected in a higher spectral noise level. The observations were taken
with the Heterodyne Array Receiver Program (HARP) 16-pixel SSB SIS receiver
\citep{Buckle2009}. The backend ACSIS correlator (Auto-Correlation Spectral
Imaging System) was configured with an operational bandwidth of 250 MHz for the
C$^{18}$O transition. $^{13}$CO was simultaneously observed with C$^{18}$O. The
resulting velocity resolution was $\sim$0.06\,km\,s$^{-1}$. The C$^{18}$O and
$^{13}$CO data were re-sampled to the velocity resolution of the $^{12}$CO data
($\sim$0.4\,km\,s$^{-1}$) taken as part of the project, in order to match
velocity bins for later outflow analysis and to improve the signal-to-noise
ratio.

The maps were taken in raster-scan mode with continuous (on-the-fly) sampling
and position switching to observe a `clean' reference position at the end of
each scan row.  The maps range in size from 5 square arcminutes (5$\times$5 arcminutes) for those
sources at distances $<$4\,kpc, to 3 square arcminutes for those between
4 and 6\,kpc. The pointing was checked with reference to a known bright molecular
source prior to each source observation. Pointing accuracy is within
$\sim$5\,arcsec, as typically expected from JCMT observations. The majority of
the baselines on each of the working receivers were flat, suitable for
detecting the weak C$^{18}$O emission. Some receivers did exhibit sinusoidal
modulations and were flagged out from the final maps accordingly. Data
reduction and display were undertaken with a custom pipeline which utilised the
\textsc{kappa, smurf, gaia} and \textsc{splat} packages which are part of the
\textsc{starlink} software maintained by the Joint Astronomy Centre
(JAC)\footnote{http://starlink.eao.hawaii.edu/starlink}. Linear baselines were fitted to
the source spectra over emission-free channels and subtracted from the data
cubes. The final C$^{18}$O cubes used in all analyses were made with a
7-arcsec spatial pixel scale. The data, originally on the corrected antenna
temperature scale \citep[$T^{*}_{A}$;][]{Kutner1981} were converted to
main-beam brightness temperature $T_{\rm mb} = T^{*}_{A}/\eta_{\rm mb}$, where
$\eta_{\rm mb} =$ 0.66 as measured by JAC during the commissioning of HARP
\citep{Buckle2009} and via ongoing planet observations. Typical spectra noise
levels are $\delta T_{\rm mb}\,\sim$\,0.8\,K in a 0.4\,km\,s$^{-1}$ bin.

\section{Mass and Radius Determination}
\label{meth_cores}

The column density and, hence, mass can be calculated if we assume LTE, and a 
constant $T_{\rm ex}$ and $\tau_{18}$ for each source, as outlined in Appendix
A. The calculations rely on the accurate determination of $\int T_{\rm mb,18}
\, d\upsilon$ over the source. The method used here is applicable to any
molecular-line tracer and source geometry. It is the combined process of
integration over velocity (i.e. creation of a moment zero map) followed by an
aperture summation over the source area. The column density and therefore mass
are calculated in each pixel via,

\begin{eqnarray}
\label{eqn6_core}
N({\rm C^{18}O}) = 5.0\, {\times}\, 10^{12} \, \frac{{\rm exp}(16.74/T_{\rm ex}) \, (T_{\rm ex} + 0.93)}{{\rm exp}(-15.80/T_{\rm ex})} \, \nonumber\\  \times \, \int T_{\rm mb,18} \, \frac{\tau_{18}}{[1 - {\rm exp} (-\tau_{18})]}\, d\upsilon\;,
\end{eqnarray}

\noindent where $N({\rm C^{18}O}) $ is in ${\rm cm^{-2}}$ and,

\begin{eqnarray}
\label{eqn7_core}
M_{\rm gas} = N({\rm C^{18}O}) \bigg[ \frac{{\rm H_2}}{\rm C^{18}O} \bigg] \mu_{\rm g}\,m(_{\rm H_2})\Omega\,D^2
\end{eqnarray}

\noindent where $\Omega$ is the solid angle of a pixel, $D$ is the distance to
the source, (H$_2$/C$^{18}$O) is the H$_2$ to C$^{18}$O abundance ratio,
where H$_2$/$^{12}$CO = 10$^{4}$ and $^{16}$O/$^{18}$O is varied according to
the relationship 58.8 $\times$ $D_{\rm gc}$(kpc) $+$ 37.1 \citep{Wilson1994},
and $\mu_{\rm g}$\,=\,1.36 is the total gas mass relative to H$_2$. A more
detailed derivation of the column density and mass are given in Appendix
\ref{AppendixA} (note the constants to change units are not included in
Equation \ref{eqn7_core} above).

\begin{figure*}
\begin{center}
\includegraphics[width=0.85\textwidth]{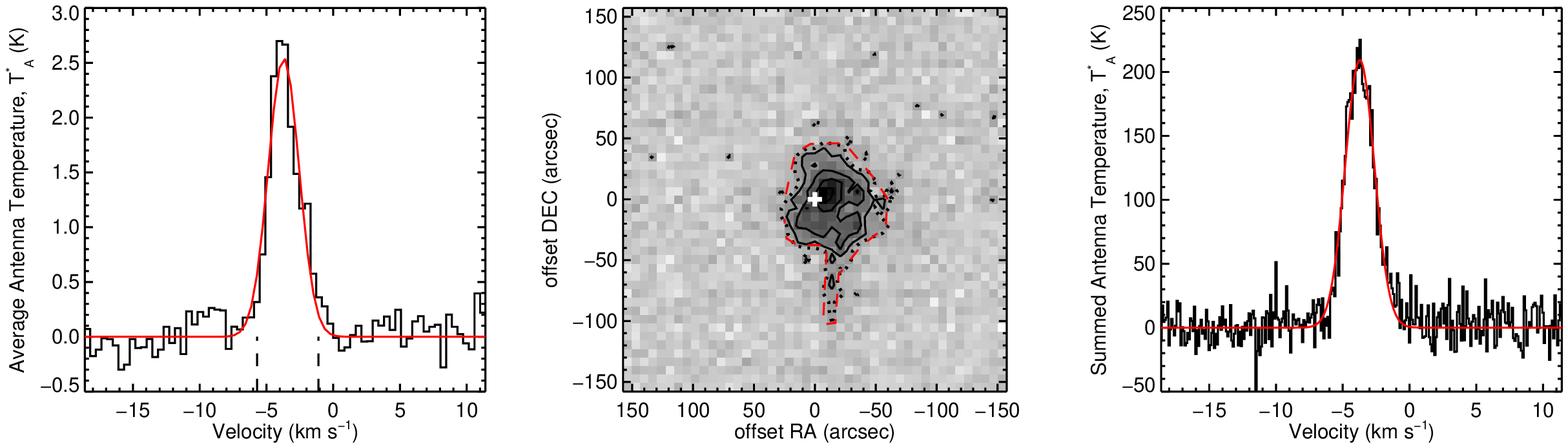}
\caption{Left: C$^{18}$O (3$-$2) line spectrum for G078.1224$+$03.6320, averaged within a 3-pixel diameter aperture centred on the RMS source location and smoothed to 0.4\,km\,s$^{-1}$. The line profile is very close to that of a Gaussian, indicated by the red line. The long dashed lines show the velocity range within which the emission is integrated. Centre: Integrated intensity map of the C$^{18}$O emission from G078.1224$+$03.6320 within the $-$5.7 to $-$1.1\,km\,s$^{-1}$ velocity range. The dotted black line indicates the 3\,$\sigma_{\rm MAP}$ contour while the solid lines are the contours at 90, 70, 50 and 30 percent of the peak emission value. The white cross indicates the RMS source location. Emission is summed within the polygon aperture, shown by the dashed red line tracing the 3\,$\sigma_{\rm MAP}$ contour level. Right: The summed C$^{18}$O spectrum extracted from the data cube within the polygon aperture defined from the integrated map (centre). The resultant spectrum has a high signal-to-noise  ratio and is Gaussian in shape (red fitted line). The velocity resolution of the summed spectrum shown is $\sim$0.1\,km\,$^{-1}$ .}
\label{fig:emiss_spec}
\end{center}
\end{figure*}

The core velocity extent (integration range) is established via a direct
investigation of the data cubes, channel by channel, specifically focusing on a
3-pixel-diameter region centred on the source location. The integration limits
are set when all emission within this region drops below 3$\sigma_{T_{\rm mb}}$
(where $\sigma_{T_{\rm mb}}$ is the standard deviation from the line-free
sections of the re-binned, $\sim$0.4-km\,s$^{-1}$ resolution spectra extracted
at every pixel) while moving away from the source $V_{\rm LSR}$ in the
directions of increasing and decreasing velocity. For the majority of sources the 3\,$\sigma_{\rm MAP}$ contour level is directly traced in order to define a polygon aperture selecting which pixels to associate with the core, from the moment zero map after velocity integration (see Figure \ref{fig:emiss_spec}, centre, where the dotted contour and red contour are 3\,$\sigma_{\rm MAP}$ and the aperture respectively). Some sources do not follow this aperture definition and are discussed below. All but five sources (the YSOs G023.6566$-$00.1273,
G094.3228$-$00.1671, G108.4714$-$02.8176 and G125.7795$+$01.7285 and the H{\sc
  ii} region G049.5531$-$00.3302) have strong C$^{18}$O emission above the
3\,$\sigma_{T_{\rm mb}}$ spectral-noise level. These 5 sources are not included
in further analysis.

Figure \ref{fig:emiss_spec} depicts the stages in the process for the source
G078.1224$+$03.6320 (IRAS 20126$+$4104, a well studied MYSO). The left panel
shows the average spectrum extracted in a 3-pixel diameter region centred on the
source, the integration ranges indicated as dashed lines, and the centre panel
is the map resulting from the integration in velocity. The mass calculated
within the defined aperture tracing the 3\,$\sigma_{\rm MAP}$ contour is
$\sim$150\,M$_{\odot}$, consistent with \citet[][who obtained 104\,M$_{\odot}$ from their
  C$^{18}$O (1$-$0) interferometric observations]{Shepherd2000} when using
their Galactocentric distance of 8.1\,kpc, a heliocentric source distance of
1.7\,kpc, a calculated temperature of 26.1\,K and without a correction made for
$\tau_{18}$.  As a final consistency check, we sum all the data within the
aperture into a single spectrum and derive a mass from a Gaussian fit to
this profile, as shown in the right panel.  The result in this case is
149 $\pm$ 6\,M$_{\odot}$, consistent with the previous estimate from velocity
integration and aperture summation. Gaussian fitting of each individual spectrum
(at each pixel) in the data cube is not used, however, as this would require
{\em a priori} knowledge of the source size and emission region, which is only
established after the integration stage.

The source radii are calculated using the area within the defined apertures.
An effective circular radius can simply be defined as 
\[ \theta_{\rm eff} = \sqrt{{area}/\pi}. \]
The deconvolved radius, assuming the JCMT has a Gaussian beam of diameter 15.3
arcseconds is

\[ \theta_{decon} =  1/2\left(\left[2 \theta_{\rm eff}\right]^2 - 15.3^2\right)^{1/2}.\]

We tested for the influence that extended, low-surface-brightness emission might have 
on these values by clipping the data below 30\,percent of the peak value
rather than at the 3\,$\sigma_{\rm MAP}$ contour.  Although the deconvolved radii
decrease, the masses also decrease by a similar factor when calculated
within the same area.  Virial-type analyses are therefore resilient against the
choice of threshold.  The mass and radius attributed to the cores are therefore
consistent and refer to the same area as in previous similar studies
\citep[cf.][]{Kauffmann2013}.

Table \ref{tab:table1} lists the sources, positions and important parameters
extracted from the RMS survey online archive. In some cases, the types are
listed as YSO/H{\sc ii} where observable characteristics are consistent with
both the MYSO and H{\sc ii}-region classification and a definitive type cannot
be ascertained \citep{Lumsden2013}. In some sources, multiple, close (a few
arcsec separation), IR-bright targets have been identified (three targets would
be listed as A, B, C in the online archive, for example). The YSO/H{\sc ii}
type is also used for these sources, if at least one MYSO and one H{\sc ii}
region is included. Individual MYSOs and H{\sc ii} regions are inseparable at
the resolution of the JCMT observations. Furthermore, where luminosities for
each target have been estimated, the total for the source is listed in Table
\ref{tab:table1}, highlighted with an asterisk (*).

\begin{table*}
\begin{center}
\caption{Source parameters for all objects in the sample, taken from the RMS survey online archive. The asterisk (*) highlights sources where multiple  IR targets have been identified. Only a small portion of the data is provided here, the full table is available in the electronic supplementary information.}
{\scriptsize
\begin{tabular}{@{}llllrrrll@{}}
\hline
MSX Source Name  & RA.   & DEC.   &  Type  &  $V_{\rm LSR}$  & Distance & Luminosity  & IRAS source  & Other  \\
                & (J2000) & (J2000) &      &   (km\,s$^{-1}$)  &  (kpc)  &  (L$_{\odot}$)  &   (offset)   & Associations \\
\hline
  G010.8411$-$02.5919 & 18:19:12 & $-$20:47:30 & YSO & 11.4 & 1.9 & 24000 & 18162$-$2048 (4$\arcsec$) & GGD27 \\
  G012.0260$-$00.0317 & 18:12:01 & $-$18:31:55 & YSO & 110.6 & 11.1 & 32000 & 18090$-$1832 (3$\arcsec$) & ...\\
  G012.9090$-$00.2607 & 18:14:39 & $-$17:52:02 & YSO & 35.8 & 2.4 & 32000 & 18117$-$1753 (11$\arcsec$) & W33A \\
  G013.6562$-$00.5997 & 18:17:24 & $-$17:22:14 & YSO & 48.0 & 4.1 &14000 & 18144$-$1723 (2$\arcsec$) & ... \\
  G017.6380$+$00.1566 & 18:22:26 & $-$13:30:12 & YSO & 22.5 & 2.2 & 100000 & 18196$-$1331 (11$\arcsec$) & ... \\
  G018.3412$+$01.7681 & 18:17:58 & $-$12:07:24 & YSO & 32.8 & 2.9 & 22000 & 18151$-$1208 (16$\arcsec$) & ... \\
  G020.7438$-$00.0952 & 18:29:17 & $-$10:52:21 & H{\sc ii} & 59.5 & 11.8 & 32000 &...& GRS G020.79$-$00.06 \\
  G020.7491$-$00.0898 & 18:29:16 & $-$10:52:01 & H{\sc ii} & 59.5 & 11.8 & 37000 &...& GRS G020.79$-$00.06 \\
  G020.7617$-$00.0638* & 18:29:12 & $-$10:50:34 & YSO/H{\sc ii} & 57.8 & 11.8 & 62000 &...& GRS G020.79$-$00.06 \\
  G023.3891$+$00.1851 & 18:33:14 & $-$08:23:57 & YSO & 75.4 & 4.5 & 24000 & 18305$-$0826 (6$\arcsec$) & GRS G023.64$+$00.14 \\
  
   \hline
\end{tabular}
}
\label{tab:table1}
\end{center}
\end{table*}

Tables \ref{tab:table2} and \ref{tab:table3} list the derived source
parameters, including temperature, optical depths, source radii and masses. In
4 of the 94 cores, the $^{13}$CO emission is optically thin $\tau_{13}\,<$1 and therefore
cannot be used reliably to estimate the excitation temperature. For these
sources, $T_{\rm ex}$ is obtained from the $^{12}$CO data obtained as part of
this project using $\tau_{12}$. These sources are flagged with an asterisk (*)
in Table \ref{tab:table2}. Although the $^{12}$CO data taken as part of this work is optically thick in all cores, these data are not used to establish the temperature of the C$^{18}$O because: the $\tau_{12}$\,=1 surface is further from the core than that of the $^{13}$CO which more closely matches the region of C$^{18}$O emission; some of the $^{12}$CO spectra also exhibit very strong self absorption; and the $^{12}$CO temperature may be elevated in some sources by the influence of outflows.

\begin{table*}
\begin{center}
\caption{Optical depths, main-beam brightness temperatures, excitation temperatures and abundance ratios for all source with strong C$^{18}$O emission. Sources with optically thin $^{13}$CO emission are flagged with an asterisk (*) to indicate that $^{12}{\rm CO}$ is used to calculate $T_{\rm ex}$. The full table is available in the electronic supplementary information.}
{\scriptsize
\begin{tabular}{@{}llllllrrr@{}}
\hline
MSX Source Name & $T_{\rm mb,13}$ & $T_{\rm mb,18}$ & $\tau_{13}$   & $\tau_{18}$   &  $T_{\rm ex}$ & Self Abs.  & $^{13}$CO/C$^{18}$O  & H$_2$/C$^{18}$O (10$^4$)  \\
\hline
 G010.8411$-$02.5919  &20.29 $\pm$ 0.21&  11.34$\pm$  0.28&  6.07   $^{+  0.35   }_{-  0.33   }$ &  0.81   $^{+  0.05   }_{-  0.04   }$ & 27.56   $^{+  0.23   }_{-  0.23     }$  &yes  &     7.4  &   419.3\\
 G012.0260$-$00.0317  & 7.78 $\pm$ 0.26&  2.42 $\pm$ 0.34 & 2.37   $^{+  0.72   }_{-  0.67   }$ &  0.33   $^{+  0.10   }_{-  0.09   }$ & 15.21   $^{+  1.37   }_{-  0.77     }$  & no  &     7.2  &   237.0\\
 G012.9090$-$00.2607  &13.26 $\pm$ 0.75&  7.49 $\pm$ 0.18 &  6.15   $^{+  0.89   }_{-  0.72   }$ &  0.83   $^{+  0.12   }_{-  0.10   }$ & 20.24   $^{+  0.82   }_{-  0.80     }$  &yes  &     7.4  &   395.8\\
 G013.6562$-$00.5997  &7.83 $\pm$ 0.30 & 3.33 $\pm$ 0.40 & 3.94   $^{+  1.00   }_{-  0.86   }$ &  0.54   $^{+  0.13   }_{-  0.12   }$ & 14.55   $^{+  0.59   }_{-  0.44     }$  &yes  &     7.3  &   301.7\\
 G017.6380$+$00.1566  &14.22 $\pm$ 0.20&  8.17$\pm$  0.28 &   6.34   $^{+  0.52   }_{-  0.48   }$ &  0.85   $^{+  0.07   }_{-  0.06   }$ & 21.25   $^{+  0.23   }_{-  0.23     }$  & no  &     7.4  &   413.4\\
 G018.3412$+$01.7681  &20.90 $\pm$ 0.24&   7.07 $\pm$ 0.33 & 2.84   $^{+  0.26   }_{-  0.26   }$ &  0.38   $^{+  0.04   }_{-  0.03   }$ & 29.47   $^{+  0.69   }_{-  0.58     }$  & no  &     7.4  &   378.1\\
 G020.7438$-$00.0952  &16.70$\pm$  0.24&  5.71 $\pm$ 0.26 & 2.85   $^{+  0.27   }_{-  0.27   }$ &  0.39   $^{+  0.03   }_{-  0.04   }$ & 24.87   $^{+  0.61   }_{-  0.52     }$  & no  &     7.3  &   325.2\\
 G020.7491$-$00.0898  &16.70 $\pm$ 0.24&  5.71$\pm$  0.26 & 2.85   $^{+  0.27   }_{-  0.27   }$ &  0.39   $^{+  0.03   }_{-  0.04   }$ & 24.87   $^{+  0.61   }_{-  0.52     }$  & no  &     7.3  &   325.2\\
 G020.7617$-$00.0638  &9.58$\pm$  0.23 & 2.60 $\pm$ 0.29  &  1.94   $^{+  0.49   }_{-  0.48   }$ &  0.26   $^{+  0.07   }_{-  0.06   }$ & 18.02   $^{+  1.68   }_{-  1.01     }$  & no  &     7.3  &   331.1\\
 G023.3891$+$00.1851  & 6.32  $\pm$0.28&  3.42 $\pm$ 0.22 & 5.67   $^{+  1.06   }_{-  0.87   }$ &  0.77   $^{+  0.15   }_{-  0.12   }$ & 12.71   $^{+  0.35   }_{-  0.33     }$  & no  &     7.3  &   307.6  \\
   
\hline  
\end{tabular}
}
\label{tab:table2}
\end{center}
\end{table*}

\begin{figure*}
\begin{center}
\includegraphics[width=0.95\textwidth]{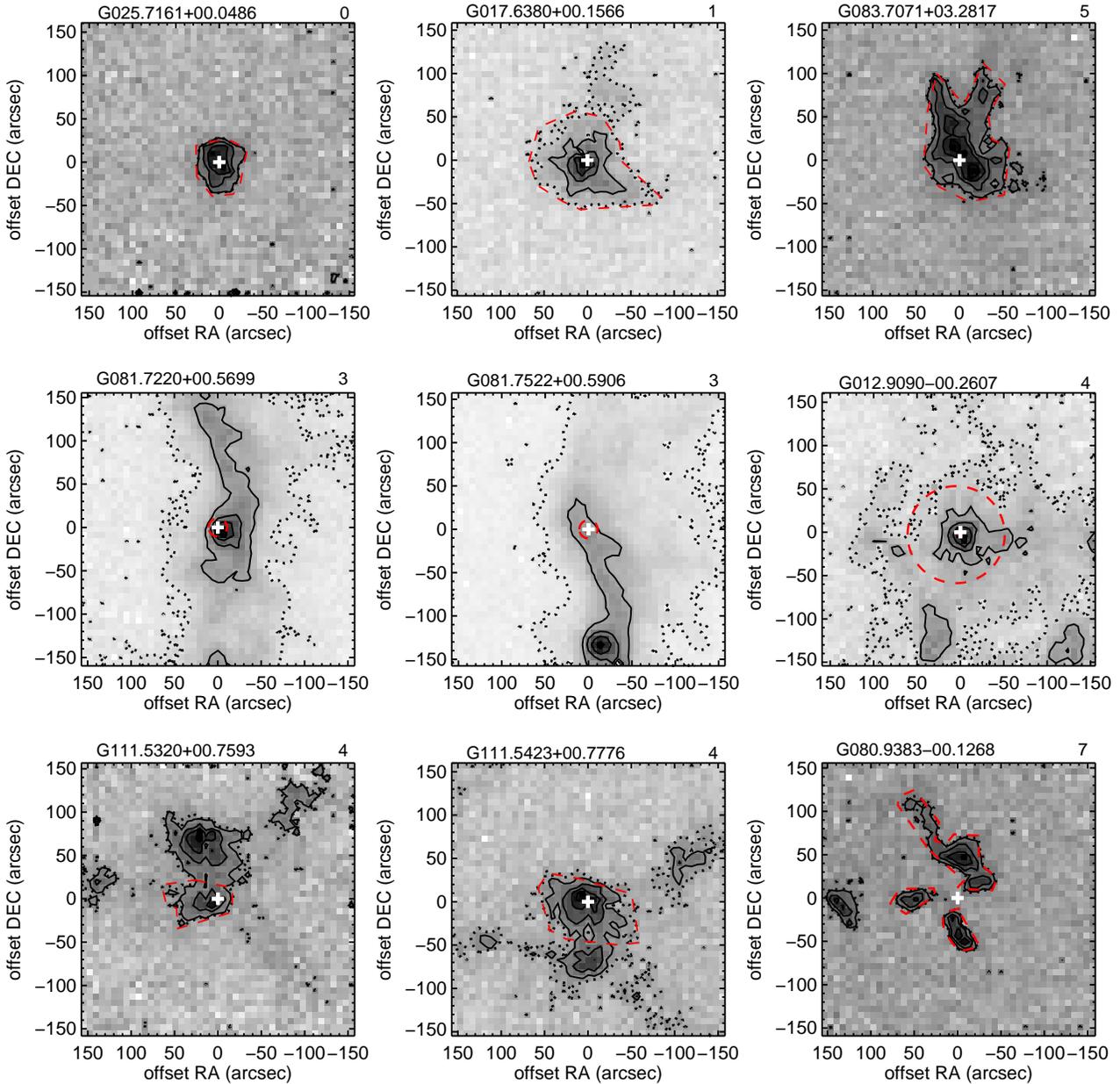}
\caption{Integrated C$^{18}$O (3$-$2) maps of sources with a range of mass flags. From left to right, top to bottom, are example sources indicating the various summation apertures for flags 0, 1, 5, 3, 4, and 7. The source name and corresponding flag is listed above the maps. Flag 3 is illustrated by two sources in the G081.7XXX region to exemplify the $\sim$3\,pixel diameter summation aperture. For flag 4, there are 3 sources shown. G012.9090$-$00.2607 has a circular aperture with a radius set from the source peak to the closest edge of the 3$\sigma_{\rm MAP}$ noise level. The sources G111.5320$+$00.7593 and G111.5423$+$00.7776 are a split structure where emission is divided mid way between the sources. For G080.9383$-$00.1268 (flag 7), it is clear that the emission is not peaked on the source location of the H{\sc ii} region. In all maps the dotted black line indicates the 3$\sigma_{\rm MAP}$ contour, while the solid contours are the emission at 90, 70, 50 and 30 percent of the peak. The white cross indicates the RMS source location and the summation aperture is indicated by the dashed red line. Flag 2 sources are not shown as they are indistinguishable from type 0 or 1 in the maps, while flag 6 sources are not shown as these can fall into any other flag category.}
\label{fig:fig2_addNEW} 
\end{center}
\end{figure*}

\begin{table*}
\begin{center} 
\caption{Measured parameters for all sources. The masses are listed from the polygon aperture method and the Gaussian summation fitting where applicable. The full table is available in the electronic supplementary information.}  
{\scriptsize
\begin{tabular}{@{}llrrrrrrrrr@{}}
\hline
MSX Source Name  & Type & Spectral  & Vel. Range  & Map Noise   & FWHM  & Centroid & Aperture  & Gaussian & Decon. & Flag  \\
                &       &Noise       &             &          &        &          & Mass  & Mass  & Radius &\\
                &    & (K)          &(km\,s$^{-1}$)&(K\,km\,s$^{-1}$) &(km\,s$^{-1}$) &(km\,s$^{-1}$) & (M$_{\odot}$) & (M$_{\odot}$) &(pc) & \\
\hline

G010.8411$-$02.5919  &     YSO  &  0.8    &  (  9.9  , 14.6   ) &  1.4  &  1.9  & 12.2  &  465 $\pm$  10  &  450 $\pm$  12&0.47  &1 \\
G012.0260$-$00.0317  &     YSO  &  1.4    &  (110.4  ,112.1   ) &  1.5  &  3.2  &110.9  &  269 $\pm$  33  &  469 $\pm$  92&0.53  &0 \\ 
G012.9090$-$00.2607  &     YSO  &  0.6    &  ( 32.6  , 40.2   ) &  1.7  &  4.4  & 36.3  & 1065 $\pm$  79  & 1104 $\pm$  26&0.64  &4 \\ 
G013.6562$-$00.5997  &     YSO  &  1.4    &  ( 46.4  , 48.9   ) &  1.6  &  3.1  & 47.8  &   108$\pm$  12  &  144 $\pm$  22&0.21  &0 \\
G017.6380$+$00.1566  &     YSO  &  0.6    &  ( 19.9  , 25.9   ) &  1.2  &  2.6  & 22.3  &  616 $\pm$  22  &  597 $\pm$  19&0.61  &1 \\
G018.3412$+$01.7681  &     YSO  &  0.7    &  ( 31.1  , 35.8   ) &  1.3  &  2.3  & 32.8  &  606 $\pm$   12 &  614 $\pm$  20&0.68  &1 \\
G020.7438$-$00.0952  &     H{\sc ii}  &  1.0    &  ( 60.0  , 60.8   ) &  0.8  &  3.3  & 59.1  &  594 $\pm$  14  & 2450 $\pm$ 150&1.06  &0 \\
G020.7491$-$00.0898  &     H{\sc ii}  &  0.8    &  ( 56.1  , 62.1   ) &  1.8  &  3.8  & 58.8  & 6067 $\pm$ 147  & 6374 $\pm$ 380&2.23  &0 \\
G020.7617$-$00.0638  & YSO/H{\sc ii}  &  1.0    &  ( 55.3  , 58.3   ) &  1.3  &  3.6  & 56.4  &  617 $\pm$  56  &  892 $\pm$ 101&0.65  &2 \\
G023.3891$+$00.1851  &     YSO  &  0.8    &  ( 73.8  , 76.7   ) &  1.0  &  2.0  & 75.3  &  358 $\pm$  37  &  376 $\pm$  30&0.49  &0 \\
\hline
\end{tabular}
}
\label{tab:table3}
\end{center}
\begin{minipage}{0.95\textwidth}
{\footnotesize
Mass flags follow the scheme:
\newline 0 $-$ {Masses calculated directly from within aperture tracing the 3\,$\sigma_{\rm MAP}$ level.}
\newline 1 $-$ {Faint filamentary structures are not included in mass calculation and are outside the aperture. In extreme cases the aperture is more circular.}
\newline 2 $-$ {Highlights cores with multiple IR-bright sources within JCMT beam (Classically flag type 0 or 1).}
\newline 3 $-$ {Source mass estimated within a 3-pixel diameter aperture (slightly over 1 beam FWHM) aperture centred on the source due to the source being part of a complex filamentary cloud complex.}
\newline 4 $-$ {Complex/multiple source regions of significant emission. Masses are split where emission peaks are separated by more than 3 pixels or are circular with a radius set as the shortest distance between the RMS source location and the 3\,$\sigma_{\rm MAP}$.}
\newline 5 $-$ {Two or more inseparable continuum cores very close within the aperture.}
\newline 6 $-$ {Luminosity estimates not from SED fitting.}
\newline 7 $-$ {Morphology suggesting that gas located at the source position has already been blown away or eroded.}
}
\end{minipage}
\end{table*}

  Not all of our source maps show an ideal source distribution, such as the test source G078.1224$+$03.6320, Figure \ref{fig:emiss_spec}. Some include diffuse emission or joined targets, thus they have a different morphology and their mass apertures do not trace the 3$\sigma_{\rm MAP}$ contour directly. In Table \ref{tab:table3} each source is given a mass flag according to its
 map morphology and aperture definition, as detailed in the table caption. These flags indicate whether the
 source masses are reliable estimates and are used in the analysis, or are
 considered unreliable, due to confusion or merging with other sources.
  Some sources are deemed to have unreliable masses where we are
   unable to define a mass aperture following the above outlined method, i.e. tracing the
   3\,$\sigma_{\rm MAP}$ contour level. Figure \ref{fig:fig2_addNEW} presents an example source from each
 flag category, except flag types 2 and 6.  The masses from sources with 0, 1
 and 2 flags are used in the analysis fully  as their mass apertures essentially follow the
   outlined method (see the Table \ref{tab:table3} caption), although flag 4 sources are used
 with caution where indicated throughout the paper. Sources with the other mass flags are not used
 in the analysis as the masses are clear underestimates, the source emission is indistinguishable from other cores in the region or emission is not associated with the source itself ( see the Table \ref{tab:table3} caption in correspondence with Figure \ref{fig:fig2_addNEW} for the list of mass flags and differences in aperture definitions). Overall there are 61 (of 89
 D\,$<$\,6\,kpc) sources flagged as 0, 1 and 2 that are regarded as having good
 mass and luminosity estimates for the cores (70 when including flag 4
 targets), these all have apertures closely following the 3$\sigma_{\rm MAP}$ contour, the remaining
 sources have different aperture definitions and are therefore not used in further analysis. The integrated images, polygon apertures and summed Gaussian spectra
 of all 99 sources (including sources with distance $>$ 6\,kpc) are presented
 in Appendix B (available online).  The sources with flags 0, 1, 2 and 4 for
 which we have derived masses are plotted in all figures henceforth (unless
 otherwise indicated).

\section{Results}
\label{samp}
\subsection{C$^{18}$O Mass and Distance}
\label{comp_dist_CO}

\begin{figure}
\begin{center}
\includegraphics[width=0.45\textwidth]{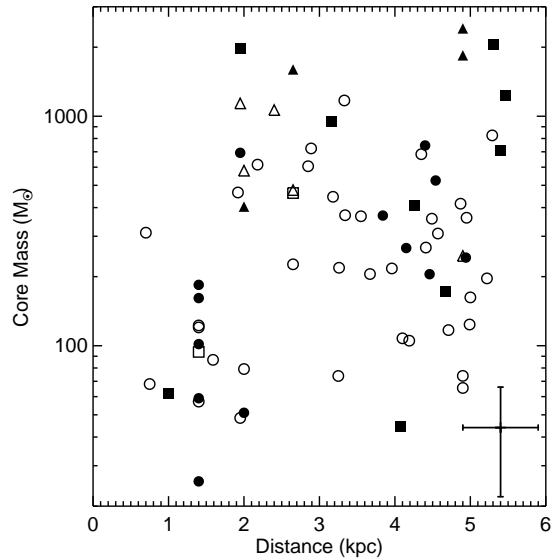}
\caption{Mass versus distance for all sources flagged as 0 and 1 (circles), 2 (squares) and 4 (triangles). The open symbols represent MYSOs and the filled ones are H{\sc ii} regions. There is a wide range of masses probed for the different source types and flags located at distances $<$6\,kpc. Note the error bars in the lower right represent a 50 percent uncertainty.}
\label{fig:mass_dist} 
\end{center}
\end{figure}

Figure \ref{fig:mass_dist} shows that a wide range of masses are sampled
at all distances in our sample, indicative of no distance dependent biases.
Furthermore, there are no biases towards MYSOs and H{\sc ii} regions
independently and the two source types span the same mass and distance ranges.
In general, the sources flagged as 4 also have masses well within the
range of those identified as having reliable estimates. It is likely that the
division of mass between sources in multi-core regions is reasonable in these
cases. There are still a few flag 4 sources where the masses
($\gtrsim$1000\,M$_{\odot}$) are likely to be overestimates due to the nature
of the aperture definitions in these cases, and may include un-associated
material with the core (see Appendix B).

All the cores are resolved at distances $<$6\,kpc.  However we cannot resolve
sub-structures within the beam.  It is likely that these cores will form
stellar clusters containing lower-mass stars as well as the targeted massive
protostar \citep[see][for an example of a relatively close region Cygnus X
  where substructure is clear]{Bontemps2010}. This is supported by the fact
that flag 2 sources are indistinguishable from flag 0 and 1 sources in the
C$^{18}$O maps, although multiple infrared sources, with separations less than
the JCMT beam, are identified in these cores.  Clearly we are studying the
properties of the natal cores of the associated star forming sites here rather
than the reservoirs associated with individual stars.

\subsection{Comparison with Continuum Masses}
\label{cont_masses_CO}

We compared the C$^{18}$O masses with those calculated from the 850\,$\mu$m
integrated fluxes from the SCUBA legacy survey \citep{DiFrancesco2008}, BOLOCAM
1.1\,mm integrated
fluxes\footnote{http://irsa.ipac.caltech.edu/data/BOLOCAM\_GPS/}
\citep{Ginsburg2013} and other 1.2\,mm observations \citep{Beuther2002b, 
Faundez2004, Hill2005}. The continuum fluxes have been used
to calculate the masses traced by cool dust (Table \ref{tab:table4}) under the
optically thin assumption via:

\begin{equation}\label{mass}
{M} = \frac{{g}\,{S}_{\nu}\,{D}^2}{ {\kappa_{\nu}}\,{B_{\nu}(T_d)} }
\end{equation} 

\noindent where ${S}_{\nu}$ is the integrated source flux, ${g}$ is the gas-to-dust ratio = 100, ${B_{\nu}(T_d)}$ is the Planck function for a black-body at a dust temperature $T_{\rm d}$ and ${D}$ is the distance to the source.  ${\kappa}_{\nu}$ is the dust opacity coefficient, calculated via ${\kappa}_{\nu}$= ${\kappa}_{0}$ ($\nu/\nu_0$)$^{\beta}$, adopting ${\kappa}_{0}$ = 1.0\,cm$^2$\,g$^{-1}$ at 250\,GHz \citep{Ossenkopf1994} and $\beta$ = 2 \citep{Beuther2002b} and hence are 1.99 and 1.19\,cm$^2$\,g$^{-1}$ for SCUBA (850\,$\mu$m) and BOLOCAM (1.1\,mm), respectively. Assuming the gas and dust are in thermal equilibrium, the calculated C$^{18}$O gas temperature for each source is used as the effective dust temperature. This is realistic given the densities of such cores \citep[$>$10$^{4}$\,cm$^{-3}$,][]{Fontani2012}. Furthermore, the mean gas temperature for all sources is $\sim$23\,K, close to typically \emph{assumed} dust temperatures \citep[e.g.][]{Hill2005} and the kinetic temperatures calculated from ammonia observations for some of these sources \citep{Urquhart2013}. The fluxes listed in the literature are used directly and thus the continuum emission regions are unlikely to be exactly matched to the emission area of C$^{18}$O and between the different continuum surveys.

\begin{figure}
\begin{center}
\includegraphics[width=0.45\textwidth]{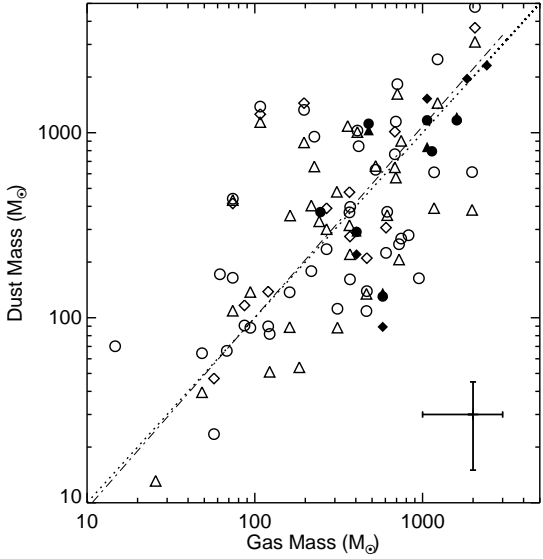}
\caption{Comparison of the C$^{18}$O masses with the 850\,$\mu$m SCUBA masses
  (circles), the 1.1\,mm BOLOCAM masses (triangles) and 1.2\,mm observations
  from various other surveys (diamonds). There is a dotted line representing
  equal masses and the dot-dashed line indicating the best bisector linear fit
  of the SCUBA and C$^{18}$O masses. These follow each other very closely. Open
  symbols represent reliable C$^{18}$O gas masses (flag 0, 1 and 2) and the
  filled symbols are for flag 4 sources (see text). The error bars in the lower
  right represent a 50\,percent uncertainty.}
\label{fig:fig4add} 
\end{center}
\end{figure}

Figure \ref{fig:fig4add} shows that the 850\,$\mu$m SCUBA, 1.1\,mm BOLOCAM and
various 1.2\,mm observations correlate very well with the C$^{18}$O masses (and
each other). The BOLOCAM masses plotted here are derived using 80-arcsec
aperture fluxes which better match the source sizes. The error bars represent a
50-percent uncertainty in the values. This is a reasonable estimate considering
the calibration accuracy of the fluxes and the potential variations in dust/gas
temperature, dust opacity, integration aperture selected and the C$^{18}$O
abundance ratio (e.g., a factor of $\sim$2.5 difference in mass can be caused
by a change in assumed temperature between 10 and 20\,K;
\citealp{Hill2005}).  There are a small number of outlier sources,
G013.6562$-$00.5997, G025.4118$+$00.1052, G030.8185$+$00.2729 and
G073.0633$+$01.7958 for which the dust-continuum masses are noticeably larger
than the C$^{18}$O masses (e.g. $>$ 5 times). Generally, the $\sim$order-of-magnitude scatter,
may mean the choice of tracer is more important for individual objects, if not for the whole sample.
However, all variations can be explained by a combination
of mismatched aperture sizes between dust and gas studies, different noise levels and
the aforementioned temperature and dust opacities.

The bisector line of best fit for SCUBA and C$^{18}$O masses follows the 1:1
line closely in Figure \ref{fig:fig4add}. The slopes for SCUBA and BOLOCAM fits
with the C$^{18}$O masses are 1.0$\pm$0.12 and 1.14$\pm$0.12 respectively. The
SCUBA observations appear to trace the same material for the majority of these
cores. For the cores with largest gas masses the BOLOCAM masses are
typically greater. However, this is likely to be due to the use of a constant aperture
size for BOLOCAM fluxes and could include more faint, extended emission of the
more massive regions in comparison to the polygon apertures chosen here.  This
is consistent with the association of C$^{18}$O with denser gas as
discussed below in Section \ref{line_rels}.

\begin{table}
\begin{center}
\caption{Continuum masses for a number of sources in the sample from BOLOCAM, SCUBA and other 1.2\,mm observation. Masses are reasonably consistent between continuum observations. Dependent upon dust opacity, temperature and observable uncertainties a generous error estimate of $\sim$50\,percent can be assumed (as Figure \ref{fig:fig4add}).}
{\scriptsize
\begin{tabular}{@{}llrrrrrrrr@{}}
\hline
MSX Source Name & Type  & BOLOCAM  & SCUBA  & 1.2 mm \\
                &       &  (M$_{\odot}$)  & (M$_{\odot}$) & (M$_{\odot}$) & \\
\hline
 G010.8411$-$02.5919  &     YSO  &  ...  &  139  &  210\\
 G012.0260$-$00.0317  &     YSO  & 2188  & 3109  & 1853\\
 G012.9090$-$00.2607  &     YSO  &  838  & 1167  & 1528\\
 G013.6562$-$00.5997  &     YSO  & 1142  & 1385  & 1259\\
 G017.6380$+$00.1566  &     YSO  &  358  &  374  &  ...\\
 G018.3412$+$01.7681  &     YSO  &  ...  &  224  &  307\\
 G020.7491$-$00.0898  &     H{\sc ii}  & 4016  &  ...  &  ...\\
 G020.7617$-$00.0638  & YSO/H{\sc ii}  & 1996  &  ...  &  ...\\
 G023.3891$+$00.1851  &     YSO  & 1087  &  ...  &  ...\\
 G023.6566$-$00.1273  &     YSO  &  584  &  ...  &  ...\\
 G023.7097$+$00.1701  &     H{\sc ii}  & 2569  &  ...  & 2966\\
 G025.4118$+$00.1052  &     YSO  &  886  & 1329  & 1446\\
 G028.2007$-$00.0494  &     H{\sc ii}  & 3099  & 4783  & 3693\\
 G028.2875$-$00.3639  &     H{\sc ii}  & 8234  & 9053  &11894\\
 G028.3046$-$00.3871  &     YSO  & 2537  &  ...  & 1377\\
 G030.1981$-$00.1691  &     YSO  &  ...  &  372  &  ...\\
 G030.6877$-$00.0729  &     H{\sc ii}  &  ...  &  ...  & 1956\\
 G030.7206$-$00.0826  &     H{\sc ii}  & 5666  &  ...  & 2308\\
 G030.8185$+$00.2729  &     YSO  &  432  &  440  &  415\\
 G033.3891$+$00.1989  &     YSO  &  357  &  ...  &  ...\\
 G037.5536$+$00.2008  &     YSO  & 4714  & 3552  & 3113\\
 G043.9956$-$00.0111  &     YSO  &  408  &  597  &  ...\\
 G045.0711$+$00.1325  &     H{\sc ii}  &  904  &  268  &  ...\\
 G048.9897$-$00.2992  & YSO/H{\sc ii}  & 1621  & 1830  &  ...\\
 G050.2213$-$00.6063  &     YSO  &  ...  &  397  &  ...\\
 G053.9584$+$00.0317  &     H{\sc ii}  &  332  &  ...  &  ...\\
 G073.0633$+$01.7958  &     YSO  &  ...  &   70  &  ...\\
 G075.7666$+$00.3424  &     YSO  &  138  &   88  &  ...\\
 G077.9550$+$00.0058  &     H{\sc ii}  &   13  &  ...  &  ...\\
 G077.9637$-$00.0075  &     H{\sc ii}  &    9  &  ...  &  ...\\
 G078.1224$+$03.6320  &     YSO  &  ...  &   90  &  138\\
 G078.8867$+$00.7087  &     YSO  &  391  &  611  &  ...\\
 G079.1272$+$02.2782  &     YSO  &  ...  &   24  &   47\\
 G079.8749$+$01.1821  &     H{\sc ii}  &   54  &  ...  &  ...\\
 G080.8624$+$00.3827  &     YSO  &   51  &   82  &  ...\\
 G080.8645$+$00.4197  &     H{\sc ii}  &   89  &  137  &  ...\\
 G080.9383$-$00.1268  &     H{\sc ii}  &   31  &   18  &  ...\\
 G081.7133$+$00.5589  &     H{\sc ii}  &  ...  &  367  &  ...\\
 G081.7220$+$00.5699  &     H{\sc ii}  &  434  &  312  &  ...\\
 G081.7522$+$00.5906  &     YSO  &  260  &  272  &  ...\\
 G085.4102$+$00.0032  & YSO/H{\sc ii}  & 1448  & 2492  &  ...\\
 G094.6028$-$01.7966  &     YSO  &  ...  &  845  &  ...\\
 G103.8744$+$01.8558  &     YSO  &  ...  &   91  &  117\\
 G105.5072$+$00.2294  &     YSO  &  480  &  ...  &  ...\\
 G105.6270$+$00.3388  &     H{\sc ii}  &  661  &  631  &  ...\\
 G109.0775$-$00.3524  &     YSO  &  404  &  179  &  ...\\
 G109.0974$-$00.3458  &     H{\sc ii}  &  220  &  161  &  275\\
 G109.8715$+$02.1156  &     YSO  &   88  &  112  &  ...\\
 G110.0931$-$00.0641  &     YSO  &  650  &  765  & 1013\\
 G110.1082$+$00.0473  &     H{\sc ii}  & 1008  & 1027  &  ...\\
 G111.2348$-$01.2385  &     YSO  &  301  &  235  &  391\\
 G111.2552$-$00.7702  &     YSO  &  316  &  371  &  477\\
 G111.5234$+$00.8004  &     YSO  &  135  &  109  &  ...\\
 G111.5320$+$00.7593  &     YSO  & 1033  & 1119  &  ...\\
 G111.5423$+$00.7776  &     H{\sc ii}  & 1215  & 1167  &  ...\\
 G111.5671$+$00.7517  &     YSO  &  657  &  952  &  ...\\
 G111.5851$+$00.7976  &     YSO  &  ...  &   12  &  ...\\
 G133.6945$+$01.2166  & YSO/H{\sc ii}  &  383  &  613  &  ...\\
 G133.7150$+$01.2155  &     YSO  &  ...  &  794  &  ...\\
 G133.9476$+$01.0648  &     H{\sc ii}  &  572  & 1148  &  ...\\
 G134.2792$+$00.8561  &     YSO  &   40  &   64  &  ...\\
 G136.3833$+$02.2666  &     YSO  &  109  &  164  &  ...\\
 G138.2957$+$01.5552  &     YSO  &  206  &  250  &  ...\\
 G139.9091$+$00.1969  & YSO/H{\sc ii}  &  ...  &  163  &  ...\\
 G141.9996$+$01.8202  &     YSO  &  ...  &   66  &  ...\\
 G192.5843$-$00.0417  &     H{\sc ii}  &  293  &  291  &  220\\
 G192.6005$-$00.0479  &     YSO  &  136  &  130  &   89\\
 G196.4542$-$01.6777  &     YSO  &  ...  &  279  &  ...\\
 G203.3166$+$02.0564  &     YSO  &   52  &   42  &  ...\\
 G207.2654$-$01.8080  & YSO/H{\sc ii}  &  ...  &  172  &  ...\\
    \hline
\end{tabular}
}
\label{tab:table4}
\end{center}
\end{table}

Table \ref{tab:table5} presents the results of a formal correlation analysis
for the gas and dust masses for different source types and flags.  Overall,
the method used to calculate the C$^{18}$O masses produces values that are
directly proportional to continuum-based mass estimates. At the resolution of
the JCMT, the C$^{18}$O and SCUBA 850-$\mu$m emission effectively traces the
same regions in the majority of the sources, and only in a few cases (where
sensitivities are notably different between the sets of data) do the emission
regions vary significantly. The BOLOCAM masses calculated from 80-arcsec
aperture fluxes also provide reasonable matches to the C$^{18}$O gas masses but
fixed apertures are not ideal for such sources. 
Continuum and C$^{18}$O line emission clearly trace the same material for these IR-bright sources.

In previous studies, depletion of C$^{18}$O is a known cause of reduced gas column density and, hence, reduced masses, for a range of cores \citep[e.g.][]{Caselli1999,Fontani2012,Yildiz2012}. Table 6 from \citet{Bergin1995} shows that the amount of CO in the gas phase varies by a factor of $\sim$2 between dust temperatures of 20 to 24\,K, which closely matches the calculated temperatures of some sources in this work. \citet{Lopez2010} also discuss how their gas masses are on average lower than those calculated from dust emission although they suggest an incorrect abundance ratio can account for this. Clearly adopting a different H$_2$/$^{12}$CO ratio other than 10$^{4}$, by a factor of 2, i.e. setting H$_2$/$^{12}$CO $\sim$ 2$\times$10$^{4}$ will already alleviate any discrepancy in dust and gas masses. Factor of 2 offsets have been found in massive, pre-stellar cores \citep{Fontani2006} and attributed to depletion. Here the reasonable correspondence of the gas and dust masses indicates that depletion of CO is not significant in the majority of these IR-bright RMS sources, especially when contrasted with the IR-dark cores in the aforementioned studies. Depletion could cause some of the scatter we see, however the already discussed variations in apertures size, dust opacity and temperature for example, can also explain this.

\begin{table}
\begin{center}
\caption{Table of Spearman rank correlation values comparing dust and gas
  masses.  The significance of a given $\rho$ value depends on the sample size
  (see Table A2.5, of \citealt{Wall2003}, from which we have also adopted the
  quoted P-values). P-values of 0.05, 0.002 and $<$0.001 represent the $\sim$2, 3 and $>$3$\sigma$ confidence levels.}  
{\footnotesize
\begin{tabular}{@{}lrrr@{}}
\hline
Correlation & Size  & $\rho$  & P-value    \\
\hline
SCUBA  &        &        &               \\
\hline
All (flag 0$+$1)        &  39 &  0.60 & $<$0.001  \\
All (flag 0, 1, 2$+$4)  &  46 &  0.60 & $<$0.001  \\ 
YSO (flag 0, 1, 2$+$4)  &  32 &  0.53 & 0.001  \\
H{\sc ii} (flag 0, 1, 2$+$4)  &  14 &  0.64 & 0.01    \\
\hline
BOLOCAM&        &        &                \\
\hline
All (flag 0$+$1)       &  33 & 0.50 &  0.002\\
All (flag 0, 1, 2$+$4) &  39 & 0.58 & $<$0.001  \\ 
YSO (flag 0, 1, 2$+$4) &  23 & 0.22 &  0.3 \\
H{\sc ii} (flag 0, 1, 2$+$4) &  16 & 0.86 & $<$0.001 \\
\hline
\end{tabular}
}
\label{tab:table5}
\end{center}
\end{table}

Finally, we stress that the masses calculated from the C$^{18}$O emission are
homogeneous in their determination and definitively associated with only the
targeted cores, given that they have a single velocity
component. \citet{Kauffmann2013} for example, include only cores where single
Gaussian components are found to avoid the arbitrary division of continuum
masses between velocity components. These C$^{18}$O observations therefore
provide an ideal way to study the core mass and its relationship to other
observables.

\begin{figure}
\begin{center}
\includegraphics[width=0.45\textwidth]{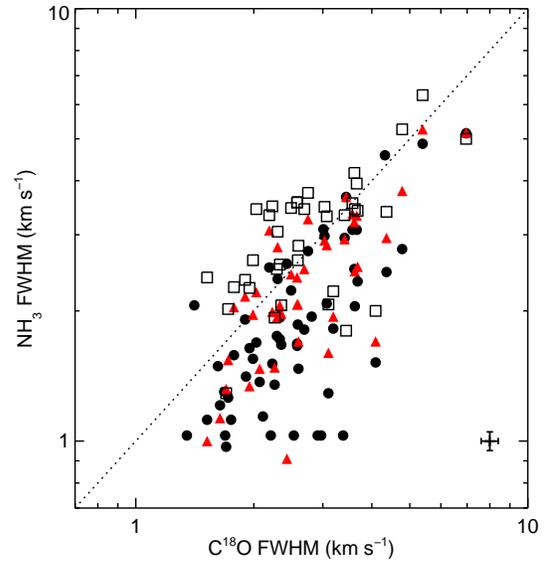}
\caption{Comparison of the C$^{18}$O linewidths with NH$_3$ (1,1) (filled
  circles), NH$_3$ (2,2) (red triangles), NH$_3$ (3,3) (squares) from
  \citet{Urquhart2011b}.The dotted
  line is the 1:1 line and is between the NH$_3$ (2,2) and NH$_3$ (3,3)transition. An average
error of $\sim$\,5\,percent from the fitting procedure is indicated.}
\label{fig:fwhmhist} 
\end{center}
\end{figure}

\subsection{Comparison with Other Linewidths}
\label{line_rels}

Previous observations of molecular gas, both in our own and other galaxies,
have led to differences in derived properties depending on the species
involved, and in particular when CO is compared to tracers of dense molecular
gas (e.g. the discussion in \citealt{Krumholz2007}).  We have compared our
C$^{18}$O linewidths ( measured using integrated spectra extracted from within the
  mass aperture regions, see Figure \ref{fig:emiss_spec} and Appendix B) with those from NH$_3$ data from the single-dish
observations of \citet{Urquhart2011b} in Figure \ref{fig:fwhmhist}.  The
increase in linewidth with excitation energy in NH$_3$ is already well reported
in similar types of sources \citep[e.g.][]{longmore2007, wienen2012}, and is
generally interpreted as due to the hotter gas being nearer the central
exciting source, and with a smaller beam filling factor in the NH$_3$
observations \citep[e.g.][]{Urquhart2011b}.  The widths are primarily driven by
motions within the gas rather than temperature broadening, which would give far
smaller values.  Our data show that the C$^{18}$O $J=3-2$ linewidths are
consistent, in magnitude, with values between the NH$_3$(2,2)
and NH$_3$(3,3) linewidths. The optically thin
C$^{18}$O 3--2 transition is tracing similar density ($\sim$\,10$^4$\,cm$^{-3}$) core
material as the ammonia line, but should trace cooler gas than both NH$_3$(2,2)
and NH$_3$(3,3), since the rotational energy levels (in temperature units) of the CO $J=3-2$ line ($\sim$33\,K) and NH$_3$(1,1) ($\sim$23\,K) are
actually better matched (cf. NH$_3$(2,2) $\sim$\,65\,K and NH$_3$(3,3) $\sim$\,125\,K).
This suggests that the CO ladder is actually
thermalised to a higher $J$ level, and that the kinematics of the $J=3-2$
transition in our sources are actually representative of that warmer gas as
well.  We also compared the line centre velocities of the NH$_3$ and C$^{18}$O,
and found they agreed well, with no deviation larger than the typical spread in the 
various NH$_3$ linewidths shown in Figure \ref{fig:fwhmhist}.  Again, this
demonstrates that both these tracers are sampling similar gas properties and
volumes.  

We also compared the $J=3-2$ C$^{18}$O and ${}^{13}$CO linewidths, with the
latter larger by at least 10\,percent.  The largest differences are seen for those
objects where the line opacity for ${}^{13}$CO is largest, so this is generally
caused by the opacity of the CO lines rather than the fact that the ${}^{13}$CO
line also traces outflow material.  A similar trend towards increasing linewidth with increasing line opacity is also seen when comparing the C$^{18}$O
$J=3-2$ data with our previous lower $J$ transition data used in deriving kinematic
distances \citep[e.g.][]{Urquhart2008}.  Indeed other similar surveys have
commented on the difference in CO linewidth for the low lying isotopologues
\citep[e.g.][]{Ao2004,Du2008,wienen2012}.  The key message here is that
different tracers may be more suitable in different circumstances.  For
example, the $J=3-2$ C$^{18}$O studied here is clearly a good tracer of
kinematics in reasonably dense clumps/cores, whereas $J=1-0$ ${}^{13}$CO is
more suitable for probing the diffuse cool gas that delineates molecular clouds as a
whole.  This also indicates why the $J=1-0$ observations are a good tracer of gas in
other galaxies, since the bulk of the mass will tend to be in the more diffuse
molecular clouds.  Notably the same may not be true in the dense environments
found in extreme star forming galaxies \citep{Harris2010}.

\begin{figure}
\begin{center}
\includegraphics[width=0.45\textwidth]{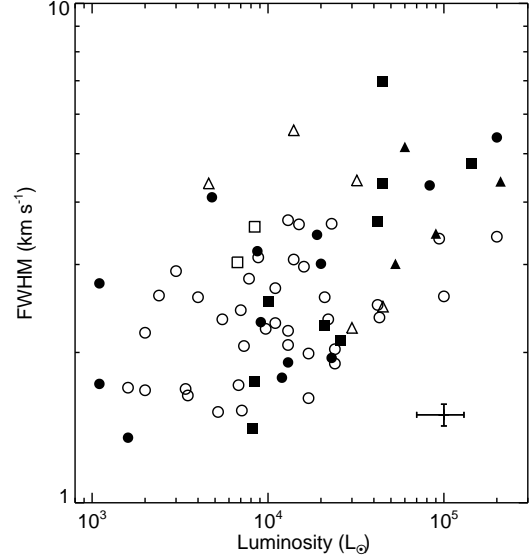}
\caption{Plot of luminosity versus C$^{18}$O FWHM for all sources flagged as 0 and 1 (circles), 2 (squares) and 4 (triangles). The open symbols represent MYSOs and the filled ones are H{\sc ii} regions respectively. A linear log-log relationship is seen between the parameters. The uncertainty from FWHM fitting is $\sim$\,5\,percent whereas that for the luminosity is a representative 30\,percent value \citep{Mottram2011a}.}
\label{fig:lumfwhm} 
\end{center}
\end{figure}

The mean FWHM values for MYSOs and H{\sc ii} regions are $\sim$\,2.6$\pm$0.1 and
$\sim$3.1$\pm$0.2\,km\,s$^{-1}$ respectively, where the uncertainties are the
standard errors (including all sources where D\,$<$\,6\,kpc and C$^{18}$O is
detected). The slightly larger FWHM for H{\sc ii} regions might be interpreted
as an evolutionary trend, where linewidths increase as the source has a greater
impact upon its surroundings. However, such an interpretation is incorrect. As
\citet{Urquhart2011b} note, any such trend is artificial and caused by the
luminosity-FWHM relationship due to the different luminosity functions for
MYSOs and fully developed H{\sc ii} regions
\citep[see][]{Mottram2011a}. Furthermore, the H{\sc ii} regions here, as
explicitly noted in Section \ref{obs_core}, were selected to be compact and
therefore should be at a similar evolutionary stage to the MYSOs. Figure
\ref{fig:lumfwhm} shows the relationship between the source luminosity and
FWHM. Evidently, the H{\sc ii} regions have a greater proportion of sources at
luminosities $>$\,10$^4$\,L$_{\odot}$ (and correspondingly larger linewidths)
which explains the slight offset of mean FWHM values reported. The Spearman
rank correlation coefficient is 0.45 at $<$0.01 significance level, interpreted
as a strong correlation for the 61 sources (with mass flags 0, 1 and 2). The relationship
does hint at the most luminous sources providing more feedback and turbulence to the
cores, possibly by driving more powerful outflows. This is examined further in
Maud et al 2015 (submitted to MNRAS).

It is worth noting that some sources (30, 32-including those
where D$>$6\,kpc) show evidence of regular velocity
gradients spatially across their cores. However, only a few of 
these appear to be aligned with, or perpendicular to the 
outflow direction.
We stress that, given the resolution of the observations, 
it is difficult to ascertain the cause of the gradients in
most cases, and whether the outflows are a major contributor.

In order to examine whether the CO lines profiles reveal the presence of infalling motions
we derived the line asymmetry parameter using the
definition of \citet{mardones1997}, 
\[ \delta \upsilon = \frac{\upsilon(thick)-\upsilon(thin)}{\Delta \upsilon(thin)} \] where $\upsilon(thick)$ and
$\upsilon(thin)$ are the velocities of the peak pixels measured from the spectra and $\Delta \upsilon(thin)$ is
the FWHM of the thin tracer. Here C$^{18}$O is the optically thin tracer (see Table \ref{tab:table2}), and the $^{12}$CO line is the optically thick one (also see Maud et al. 2015 submitted to MNRAS). The optically thin FWHM
values of C$^{18}$O are those given in Table \ref{tab:table3}. There
is a spread of asymmetry values, ranging from $-$1 to +1.5 (where by the
\citealt{mardones1997} criterion $|\delta \upsilon|>0.25$ indicates asymmetry).  The
results show that 23 sources have a red asymmetry, and 15 a blue asymmetry.
At face value this can be taken as a lack of evidence for ongoing infall in our
sample.  However, we note that \citet{fuller2005} also find a similar result in
the profiles of the transitions they study which have similar excitation temperature and
critical density, even though the same sources show strong blue asymmetry in,
for example, HCO$^+$ $J=1-0$.  Our results appear to add evidence to their
comment that it is the dense gas tracers that are best tracers of large scale
infall. CO appears to be a poor tracer of infall for these cores, while
HCO$^{+}$ (4$-$3) is usually much better suited to these size scales \citep{Klaassen2007}.

\section{Analysis}
\label{analysis}
\subsection{Correlations within the Observables}
\label{correlations}
We examined possible correlations amongst the observable parameters present in
our sample, and within the wider information held as part of the RMS database.
In addition we also considered combinations of these parameters.  In particular
we used the derived values from the C$^{18}$O observations for radius,
$T_{\rm ex}$, $\tau_{13}$, mass, line asymmetry and FWHM as well as
combinations of these such as gas surface density, as discussed below.  We took
information on the luminosity, infrared colours and galactocentric radius,
$R_{\rm gc}$  from
the database \citep{Lumsden2013}, as well as a selection of the NH$_3$
properties from \citet{Urquhart2011b}.  We used a Kendall correlation method
for all of these comparisons. The Kendall
method is generally held to be better in the presence of errors in the data \citep{Wall2003},
however typically both the Kendall and Spearman rank correlations lead
to the same inferences. Here we choose the Kendall correlation since some
parameters have ties in the data, where both
pairs have the same ordinal.

The net result  is that there are relatively few strong correlations,
and many of those that are present are driven by the one obvious dominant
relation, that between mass and radius.  A partial correlation analysis confirms
this.  The only other strongly significant parameters are luminosity and FWHM.
Table \ref{tab:kendall} shows the resultant correlations that have significance
values $<0.002$.  

The correlations that are not dependent on mass, luminosity, radius or FWHM can
be summarised briefly as follows:
\begin{itemize}
\item {The NH$_3$ kinetic temperatures, $T_{\rm kin}$, 
  from \citet{Urquhart2011b} agree reasonably with the $T_{\rm ex}$ values we derive here (See Appendix A).}
\item {The kinetic temperatures are correlated with both infrared colour measures we
use, $F_{W3}/F_K$ and $F_8/F_K$, with $T_{\rm ex}$ showing a slightly weaker
correlation than $T_{\rm kin}$ (formally not significant at our threshold level for $F_8/F_K$).}
\item {$\tau_{13}$ is anticorrelated with $R_{\rm gc}$.}
\end{itemize}
The first of these essentially shows that the CO
and NH$_3$ are tracing similar material, as we argued previously in Section
\ref{line_rels}.  The second is curious since we might expect the redder
objects to be more embedded and have lower kinetic temperatures rather than
higher. It may indicate that these warmer sources simply have a more centrally
concentrated mass distribution.  This is not something that we can test with
the spatial resolution of the current data.  Finally the anticorrelation
between $\tau_{13}$ and $R_{gc}$ may simply be a reflection of the observed
anti-correlation between luminosity and $R_{\rm gc}$ found by \citet{Lumsden2013}
in the full RMS sample (since $\tau_{13}$ is correlated with density, which is
weakly correlated at the P$=0.025$ significance level with luminosity), even
though no significant correlation between these variables is found in the much
smaller sample here.

\begin{table}
\begin{center}
\caption{Table of Kendall rank correlation values for various derived
  observables from both this paper and the wider RMS database.  $R$ is the
  deconvolved source
  radius, $M$ the mass, $L$ the luminosity, $\Delta \upsilon$ the FWHM, $\rho$ the
  density, $\Sigma$ the gas surface density, $\tau_{13}$ the optical depth of
  $^{13}$CO, $R_{\rm gc}$ the Galactocentric radius, $t_{\rm ff}$ the free-fall time,
  $T_{\rm ex}(\rm CO)$ the excitation temperature of the $^{13}$CO, $T_{\rm kin}({\rm
    NH}_3)$  the kinetic temperatures derived from NH$_3$ by
  \citet{Urquhart2011b}, $F_8/F_K$ the ratio of 8$\mu$m MSX flux with $K$ band
  flux as derived for \citet{Lumsden2013} and  $F_{W3}/F_K$ the ratio of the
  WISE band 3 flux (approximately $10\mu$m) with $K$ band
  flux.  The $-$ indicates data where there are ties so no well defined
  significance exists. The significance is the probability that the null
  hypothesis, i.e. there is no correlation, is correct.
}  
{\footnotesize
\begin{tabular}{@{}lrrr@{}}
\hline
Correlation & Sample Size  & $\tau$  & Significance    \\
\hline
$R-M$ & 58 & 0.78 & $<0.0001$ \\
$R-\Delta \upsilon$ & 58 & 0.29 & $0.002$ \\
$R-L$ & 58 & 0.47 & $<0.0001$ \\
$M-\Delta \upsilon$ & 58 & 0.39 & $<0.0001$ \\
$M-L$ & 58 & 0.51 & $<0.0001$ \\
$\Delta \upsilon-\rho$ & 58 & 0.35 & $0.0001$ \\
$\Delta \upsilon-\Sigma$ & 58 & 0.35 & $0.0001$ \\
$\tau_{13}-\rho$ & 58 & 0.35 & $0.0001$ \\
$\tau_{13}-R_{\rm gc}$ & 58 & $-0.37$ & $<0.0001$ \\
$\tau_{13}-\Sigma$ & 58 & 0.35 & $0.0001$ \\
$t_{\rm ff}-\Sigma$ & 58 & $-0.42$ & $<0.0001$ \\
$T_{\rm ex}(CO)-T_{\rm kin}({\rm NH}_3)$ & 41 & 0.52 & - \\
$T_{\rm kin}({\rm NH}_3)-F_8/F_K$ & 37 & 0.42 & - \\
$T_{\rm kin}({\rm NH}_3)-F_{W3}/F_K$ & 31 & 0.41 & - \\
$\Sigma_{\rm SFR}=\Sigma/t_{\rm ff}$ & 58 & 0.49 & $<0.0001$ \\
\hline
\end{tabular}
}
\label{tab:kendall}
\end{center}
\end{table}

Mass, luminosity, size of clump and FWHM are all positively correlated with
each other (the luminosity-FWHM correlation has a significance level of only
0.003 however).  These four collectively are what is expected from the scaling
relationships of \citet{Larson1981}.  Indeed, appropriate projections of the
mass-luminosity-radius plane, for example, result in most data points being
strongly clustered (i.e., there is a mass-radius-luminosity fundamental plane),
though the luminosity component of this relationship is small, and the same is
true for mass, FWHM and radius.  The first of these can be understood, for
example, as Larson's third relation (mass-radius) but modified for the case of cores
where a not insignificant fraction of the mass is now locked into stars (and
hence not traced by the gas mass: cf. Section \ref{sfe}).  The weak luminosity
dependence of this mass-radius relation effectively compensates by directly
tracing the mass already locked into stars.  The strong correlation of mass and
luminosity reflects the fact that these sources are of a similar evolutionary
state, otherwise more scatter would be evident.  \citet{Davies2011} shows that
H{\sc ii} regions can be of a similar age to massive protostars if the central source
is itself more massive (i.e. more massive objects evolve more quickly towards the
zero age main sequence).  This conclusion is justified further in Section
\ref{comp_lum_CO}.  It is notable that the FWHM and radius (Larson's first
relation) gives the weakest correlation, as shown in Table \ref{tab:kendall}.
We will discuss this and the other Larson relations in detail in the following
sections, as well as comparing our results with those of \citet{Urquhart2014b}
and \citet{Heyer2009}.  \citet{Heyer2009} used $^{13}$CO $J=1-0$ data from
the Galactic Ring Survey \citep{jackson2006} to re-examine the underlying
physical principles that lead to Larson's scaling relations for molecular-cloud-sized 
structures and, hence, sampled much larger regions but with much lower
surface densities.  \citet{Urquhart2014b} analysed ATLASGAL sources (including
those with RMS counterparts) and hence studied the continuum dust emission of similar
regions to us.

Some of the properties that show no correlation are also worth
mentioning.  The line asymmetry parameter shows no correlation with any other
properties, including the line opacity itself for example.  There are no
correlations between infrared colour and properties such as mass, or line
asymmetry, where both red and blue asymmetric data show the same average
colour, or between most of the CO properties (including opacity and column
density) and infrared colour.  The beamsize of our JCMT data are not dissimilar
to many of the mid- and far-infrared data we use the infrared colours from.  The
natural explanation for the lack of correlation therefore is that there is
substructure within the beam for both sets of data.  In particular the CO data
presented here traces cooler gas, whereas the infrared colours we have are
predominantly a combination of hot and warm dust and extinction.  The spatial
scales of these components should differ considerably but we are unable to
probe such detail.  This also tallies with the discussion above regarding the actual
correlation seen between $T_{\rm ex}$ or $T_{\rm kin}$ and colour.

\subsection{Mass and Radius}
\label{comp_rad_CO}
\begin{figure}
\begin{center}
\includegraphics[width=0.45\textwidth]{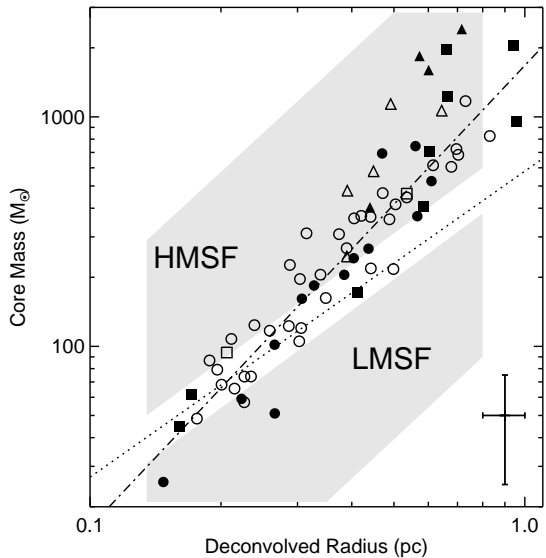}
\caption{Core mass versus the deconvolved effective radius assuming spherical cores. Open and filled symbols are MYSOs and H{\sc ii} regions respectively. The sources with 0 and 1 flags are represented by circles, flag 2 by squares and flag 4 sources by triangles. The general linear fit trend follows a power law ($\gamma$ = 2.0 $\pm$ 0.1) indicated by the dot-dashed line. The dotted line is that from \citet{Kauffmann2010a} where massive star forming regions exceed a mass of 580\,$R_{\rm pc}^{4/3}$. The upper and lower light grey boxed regions represent the high and low mass star forming regions presented in \citet{Kauffmann2010a} after a reduction in mass by 1.5 to correct for the difference in dust opacity used.  Note the larger, more massive ATLASGAL sources from \citet{Urquhart2014b} are predominantly located in the HMSF region where M$\gtrsim$1000\,M$_{\odot}$ and R$\gtrsim$0.5\,pc. A representative uncertainty of 50\,percent in mass and an 10\,percent illustrative uncertainty for the radius is indicated to the lower right.}
\label{fig:mass_rad} 
\end{center}
\end{figure}

Previous observations of molecular clouds/cores suggest a power-law scaling
between mass and length scale (size, radius) of the form $m\propto r^{\gamma}$,
where $\gamma \sim 2$
\citep[e.g.][]{Larson1981,Elmegreen1996,Kirk2013,Kauffmann2013}. Their
observations include different regions, associations (low- and high-mass stars)
and observations using both molecular line tracers and continuum
emission. Figure \ref{fig:mass_rad} shows the C$^{18}$O masses
plotted against deconvolved radius for our sample. The linear trend when
including the flag 0, 1 and 2 sources follows the power law $m \propto r^{2.0
  \pm 0.1}$, where the uncertainty only accounts for the spread in the raw mass
and radius values. Note, exchanging the deconvolved radii for the effective
radii has a minimal effect and slightly steepens the slope to $m \propto r^{2.1
  \pm 0.1}$.

The ATLASGAL clumps associated with RMS sources investigated by
\citet{Urquhart2014b} span a larger range of radii and mass, and follow a
slightly shallower slope ($\gamma\sim1.75\pm0.04$) overall. There is a turnover
in their data at higher masses, such that if we restrict the range to that of our sample
a slightly steeper slope $\sim$1.9 would be recovered. However, the
discrepancy in slopes can be attributed to the different methods in which mass and radius
are calculated (we see that the slope changes when using effective
radius). \citet{Urquhart2014b} find that the large scale clump properties
follow the same trend from embedded maser sources to extended H{\sc ii}
regions (i.e. less to more evolved). They suggest the clump properties must therefore
be set prior to the onset of
star formation and that the subsequent evolution of massive protostars, and
their feedback, does not effect the clumps overall.  Note that the internal
structures may still evolve (cf. the study of \citealp{Kauffmann2010b} which
examined mass as a function of radius within single clumps, as opposed to the
inter-clump comparisons given here).  We basically find the same relations
as \citet{Urquhart2014b} from the gas as opposed to dust.  \citet{Heyer2009}
find $\gamma\sim2.17\pm0.08$, again in reasonable agreement with the slope
found here.

Where the Heyer data differs from the smaller clump/core sized regions
studied here is in the offset of this relationship, which lies at lower masses in the Heyer et al sample.
Fundamentally, both our
sources and those of \citet{Urquhart2014b} are significantly more massive than the
similar size regions in the sample of \citet{Heyer2009}.
 It is worth reiterating that Heyer et al. used $J=1-$0 $^{13}$CO data. The comparison in Section \ref{line_rels} for our objects
  suggests such lines should be broader (due to opacity) than the $J=3-$2 C$^{18}$O.
  Therefore, if anything, the {\em true} offset from a fair comparison with the Heyer et al. data may be even
  lower than suggested by the raw data. This is contrary to
the spirit of the relation that Larson initially proposed.

\begin{figure}
\begin{center}
\includegraphics[width=0.45\textwidth]{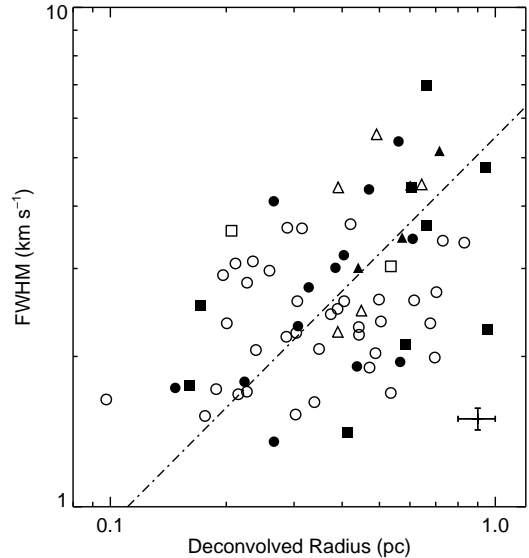}
\caption{Deconvolved source radius versus C$^{18}$O FWHM for MYSOs
  (open) and H{\sc ii} (filled) regions flagged as 0 and 1 (circles), 2
  (squares) and 4 (triangles). The bisector fit best representing the data is
  shown as the dot-dashed line, and obeys FWHM $\propto r^{0.8\pm0.5}$. An average
  uncertainty of 5\,percent in FWHM and an illustrative 10\,percent uncertainty in
radius is indicated to the lower right.}
\label{fig:fwhm_rad} 
\end{center}
\end{figure}

\subsection{FWHM-Mass and FWHM-Radius relationships}   
\label{mass_fwhm}

Figure \ref{fig:fwhm_rad} shows the log-log relationship between the C$^{18}$O
FWHM and deconvolved source radius. This comparison is analogous to the type-2,
single tracer multi-core relationships presented in \citet{Goodman1998} where
the FWHM is observed to decrease for smaller cores. The meaningful result of
such a relationship is to examine whether the cores are virialised, where
FWHM\,$\propto r^{0.5}$ is expected. The bisector fit best represents the data
given the scatter \citep[compared with ordinary least squares fitting, OLS,
  see][]{Isobe1990}, resulting in FWHM\,$\propto$r$^{0.8\pm0.5}$, roughly
consistent with virialised cores.  However, as already noted (Table \ref{tab:kendall}),
the significance of the correlation seen here is very weak (it is notably
stronger for the data from \citealt{Heyer2009}).  As with the
results of the previous section, a noticeable offset exists in the
relationship seen here and that of molecular cloud scales from the data of
\citet{Heyer2009}.  The cores here exhibit much larger FWHM values.  One
possible cause for the weaker correlation and larger values is that on these
scales feedback has a more significant impact compared to gravitational motions.

\begin{figure}
\begin{center}
\includegraphics[width=0.45\textwidth]{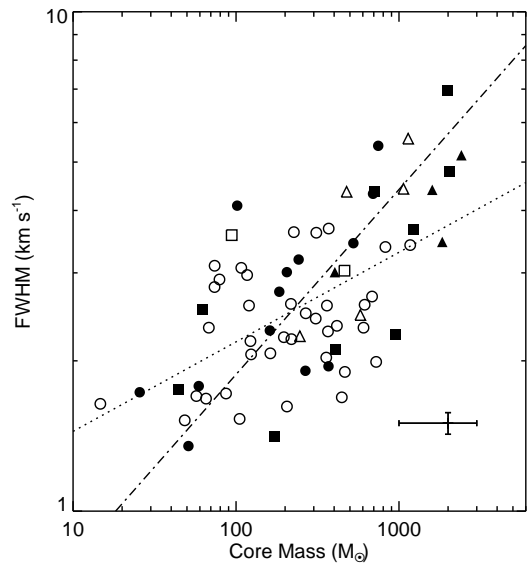}
\caption{Core mass versus the FWHM linewidth for MYSOs and H{\sc ii} regions (open and filled symbols respectively). Flag 0 and 1, 2 and 4 are represented by circles, squares and triangles. The plotted dotted line is the OLS best fit with a similar slope (FWHM $\propto m^{0.18\pm0.04}$) to the by-eye fit from \citet{Larson1981}. The bisector fit better represents the data given the scatter and has a steeper slope, $FWHM \propto m^{0.37\pm0.25}$. The average uncertainty of 5\,percent in FWHM and representative 50\,percent uncertainty for the mass is shown to the bottom right.}
\label{fig:fwhmmass} 
\end{center}
\end{figure}

\citet{Larson1981} noted a clear relationship between mass and velocity
dispersion (linewidth) over many orders of magnitude in mass.  Recent studies
have confirmed this, though the tight relationship Larson found is less well
reproduced.  As the core masses are strongly related to the source radii, which
in turn are weakly correlated to the linewidth, there is expected to be a link
between mass and FWHMs for these sources.  Figure \ref{fig:fwhmmass} shows the
correlation present between the mass and FWHM linewidths.  A bisector best fit
to the data indicates FWHM $\propto m^{0.37\pm0.25}$, although the OLS fit is
noticeably shallower, FWHM $\propto m^{0.18\pm0.04}$, and closer to that found
by \citet{Larson1981} who fit by eye. The slopes are consistent given the
uncertainties as a result of the scatter in the data.  The \citet{Heyer2009}
data give a similar slope, $\propto m^{0.20\pm0.02}$, but again the
relationship is offset, with smaller linewidths for molecular-cloud-sized
regions at the same masses as in our sample.

\subsection{Virial Mass and Gas Surface Density}   
\label{comp_virial}

\begin{figure}
\begin{center}
\includegraphics[width=0.45\textwidth]{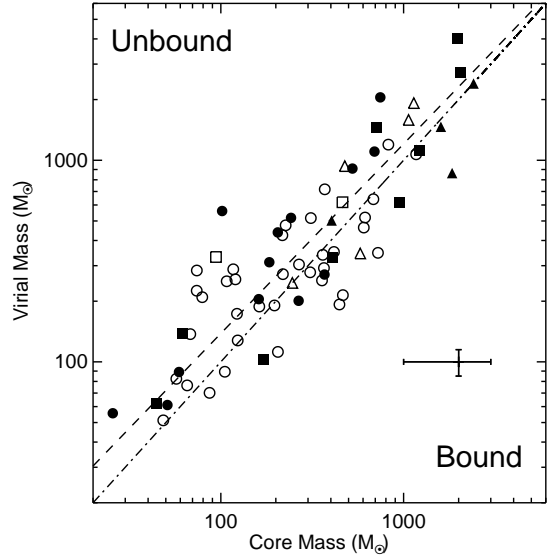}
\caption{Core mass versus the virial mass for all flag 0, 1 (circles), 2
  (squares) and 4 (triangles) cores. The open and filled symbols represent
  MYSOs and H{\sc ii} regions, respectively. The plotted dot-dashed line is
  that of equal mass, while the dashed line indicates the bisector best
  fit. Most sources have $M_{\rm core}$ $\sim$ $M_{\rm vir}$ and are $\sim$virialised,
  with a weak tendency for the lower mass cores to be less so than high mass
  cores. A 50\,percent representative uncertainty in core mass and an estimated
  uncertainty of 15\,percent in Virial mass (propagated from radius and FWHM uncertainties)
are indicated to the bottom right.}
\label{fig:virial} 
\end{center}
\end{figure}

Star formation has already occurred in these cores as they harbour at least one
IR-bright protostar. However, the virial masses can also be used as an additional test to
investigate the impact of feedback from these sources. Figure \ref{fig:virial}
shows the C$^{18}$O core masses against the calculated virial masses following
\citet{Maclaren1988}:

\begin{equation}\label{virial}
{M_{vir}}({\rm {M}_{\odot}}) = 126 R ({\rm pc}) \; FWHM^2 ({\rm km\,s^{-1}}),
\end{equation} 

\noindent where $R$ is the deconvolved radius, we use the FWHM of the
C$^{18}$O emission and assume spherical cores with a $\rho \propto r^{-2}$
density distribution and no magnetic support.  Changes in geometry, or the density
law (Figure \ref{fig:mass_rad} would imply a $\rho \propto r^{-1}$ for a
spherical geometry for example), generally increases the virial mass by up to
50\,percent (cf. \citealt{Kauffmann2013}).  Furthermore, we note that the virial masses
would have artificially been elevated if optically thick tracers such as
$^{12}$CO and $^{13}$CO were used for this analysis given their larger
linewidths.  Using the $^{13}$CO (3$-$2) data would result in an increased
virial mass by a factor of $\sim$2, furthermore optically thick, lower density
$J$ =1$-$0 transitions could cause an increase larger than a factor of $\sim$5 if
linewidths are more than double those measured here. It is therefore important
that confirmed, optically thin tracers are used when calculating virial masses.

\begin{figure}
\begin{center}
\includegraphics[width=0.45\textwidth]{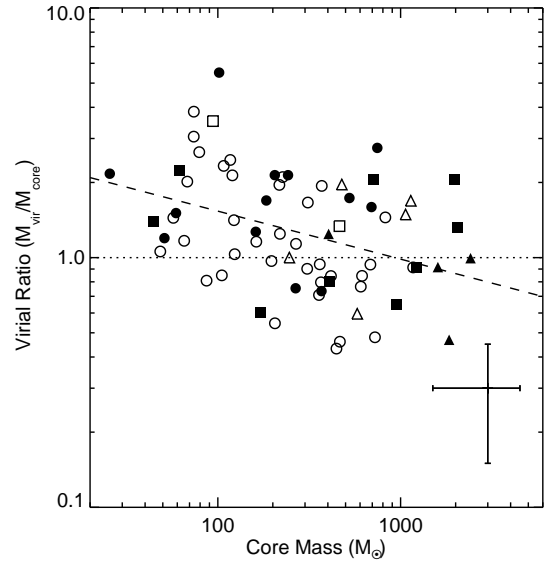}
\caption{Virial ratio, $M_{\rm vir}$/$M_{\rm core}$ versus the core mass itself for flag 0, 1 (circles), 2 (squares) and 4 (triangles) cores. It is clear to see the more massive cores are closer to equilibrium, whereas the smaller ones have larger virial masses than the core mass. Representative 50\,percent errors in the core
mass and virial ratio (dominated by core mass uncertainty) are indicated to the bottom left.}
\label{fig:virial_ratio} 
\end{center}
\end{figure}

The C$^{18}$O gas masses in the sample are closely matched with the virial
masses and are distributed about the 1:1 line (Figure \ref{fig:virial}
dot-dashed line), the results are consistent with those found for a smaller
sample of IR-bright sources by \citet{Lopez2010}. However, the bisector fit to
the data shows that the lower mass cores are actually skewed towards being unbound
whereas the more massive ones are $\sim$virialised, more clearly shown in
Figure \ref{fig:virial_ratio} plotting the virtial ratio versus core mass.
This trend is seen by (\citealt{Urquhart2014}, and
\citealt{Kauffmann2013} for individual datasets) where $M_{\rm vir}$/$M_{\rm core}$
increases with decreasing clump mass.  \citet{Kauffmann2013} argue that these
trends can be explained if higher mass cores collapse and evolve rapidly
through to the formation of their final stars, whereas lower mass cores may
still have support present from, for example, outflow activity.  Our results
here are consistent with this picture.  Only higher spatial resolution data
will finally allow us to determine whether these global trends are reflected
for individual protostellar sites, and how these interact with each other as a
core collapses as a whole.

There is no correlation between the surface density and radius, as expected for
the Larson-like mass-radius scaling we observe in our data.  There is however a
scatter of almost an order of magnitude in the surface density.  This
correlates positively with the opacity in the $^{13}$CO line, in the sense
that the highest surface densities have the highest opacities. However this
can also be viewed in terms of the discussion regarding whether the Larson
relations are actually a function of the limiting surface density (or column
density, opacity or extinction) seen for molecular cloud samples rather than an
underlying self-similar scaling relationship
\citep[e.g.][]{Lada2010,Lombardi2010,heiderman}, though see \citet{Burkert2013}
for a discussion of possible biases in the actual observational evidence for
this.  The rationale is that below this threshold there is insufficient
shielding to allow molecular gas to form efficiently (see also the discussion in
\citealt{Evans2014}).  The limiting star-formation surface density
found by, e.g. \citet{heiderman}, lies below the threshold of our sample as
expected given Figure \ref{fig:mass_rad}.  We see no trend in the star-formation-rate 
surface density with gas surface density (as is evident from
Mass-Luminosity plots, which essentially show the same data given the correct
mass-radius correlation). However the correlation of surface density with
opacity is consistent with this general concept of a threshold.

\begin{figure}
\begin{center}
\includegraphics[width=0.45\textwidth]{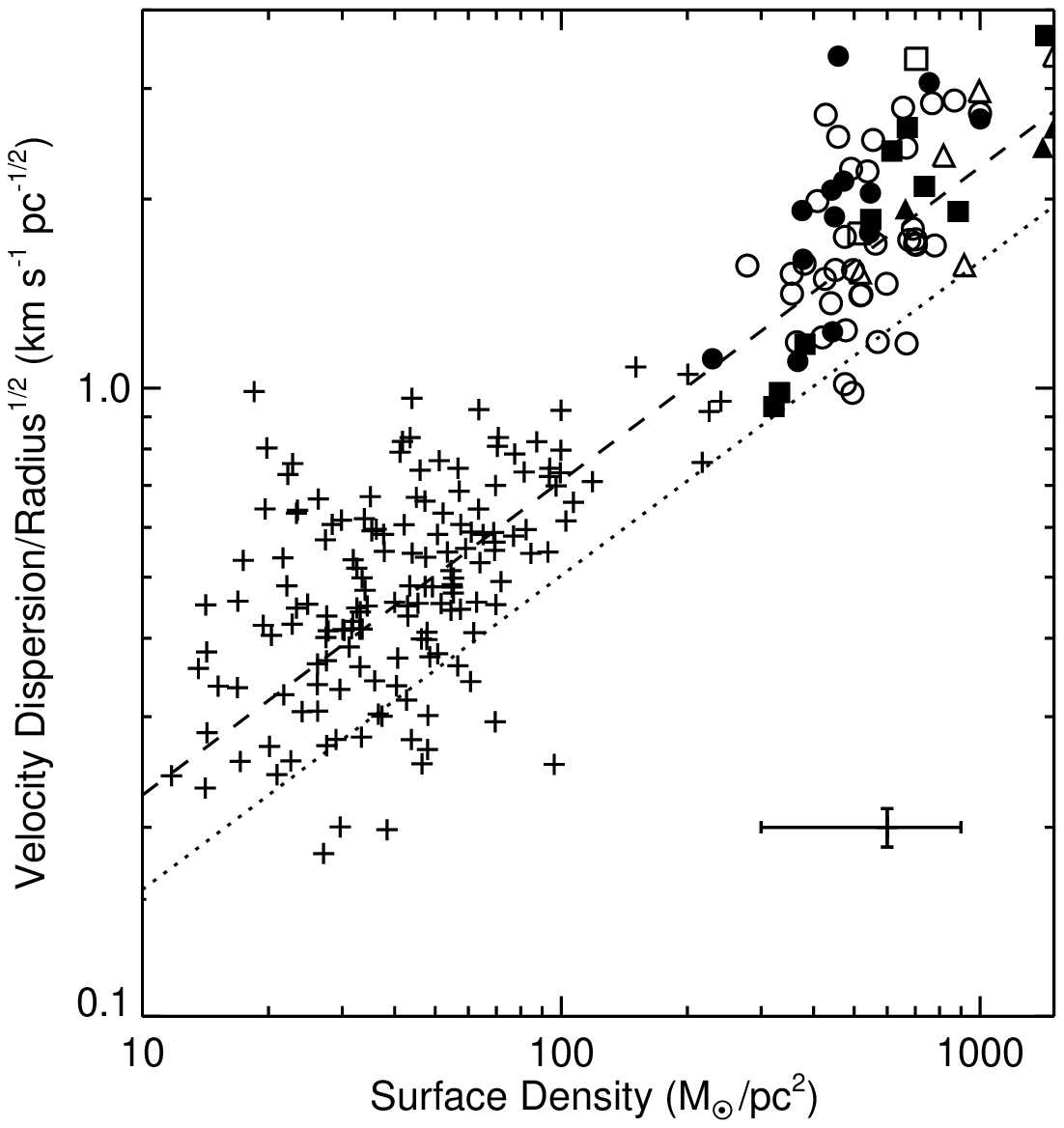}
\caption{The virial-surface density relation from \citet{Heyer2009}, for flag 0
  and 1 (circles), 2 (squares) and 4 (triangles) cores.  The crosses are
  taken from \citet{Heyer2009}.  The dotted line is that expected for
  virialised bound clouds, and the dashed line that for clouds in free-fall as
  outlined in \citet{ballesteros2011a}. The propagated uncertainties are $\sim$\,55\,percent and
  $\sim$\,7\,percent for surface density and velocity dispersion/radius$^{1/2}$
respectively. }
\label{fig:heyer} 
\end{center}
\end{figure}

\citet{Heyer2009} pointed out that, if virial motions were the primary drivers
of the Larson relationships, thus leading to a single surface density value
regardless of size of cloud, then we should expect to see the ratio
$\sigma_v/\sqrt{R}$ as a constant.  Figure \ref{fig:heyer} shows the equivalent
plot for our data.  This clearly shows that this scaling depends on surface
density, just as they found (and as seen in many other samples including that
of Larson $-$ \citealt{ballesteros2011a}).  In part this is consistent with the
other results we have already discussed.  The {\em scaling} apparent in
Larson's relations depends critically on both the objects observed (i.e. real
physical differences) and on the methods used.  Crucially, the area over which
the surface density is calculated should be matched with the appropriate
kinematic data, and not cross matched with other measures, and as we have
already noted, derived from an optically thin tracer that is representative of
the structures in question.  In principle if these simple guidelines are
followed we can compare relatively dissimilar samples.  Figure \ref{fig:heyer}
shows the sample of clouds from \citet{Heyer2009}, as well as our own.
\citet{ballesteros2011a} also showed a similar plot for data from the infrared dark
cloud sample of \citet{gibson2009}.  

We show both the virial line expected for this plot from \citet{Heyer2009} and
the free-fall prediction from \citet{ballesteros2011a}.  The data are best
matched by the latter.  This relationship is essentially carrying the same
messages as the fact that mass, radius and FWHM form a fundamental plane, with
the large velocity dispersion objects being those which are most massive for a
particular radius.  It also essentially embeds the same result seen in Figure
\ref{fig:virial_ratio}, since lower virial ratios for more massive objects tend
to tally with higher velocity dispersions for higher surface densities as well.

Finally we note that we find a strong linear correlation between the star
formation rate surface density (derived from the luminosity \citep{Kennicutt1998})
and the gas surface density ratioed with the free
fall time.  This is in agreement with the discussion in \citet{Krumholz2012},
but we would note that the correlation seen is inevitable given the form of the
mass radius relationship.

\section{Star Formation Efficiency and Evolution} 
\label{comp_lum_CO}

\subsection{Gas Mass and Luminosity}
\begin{figure*}
\begin{center}
\includegraphics[width=0.95\textwidth]{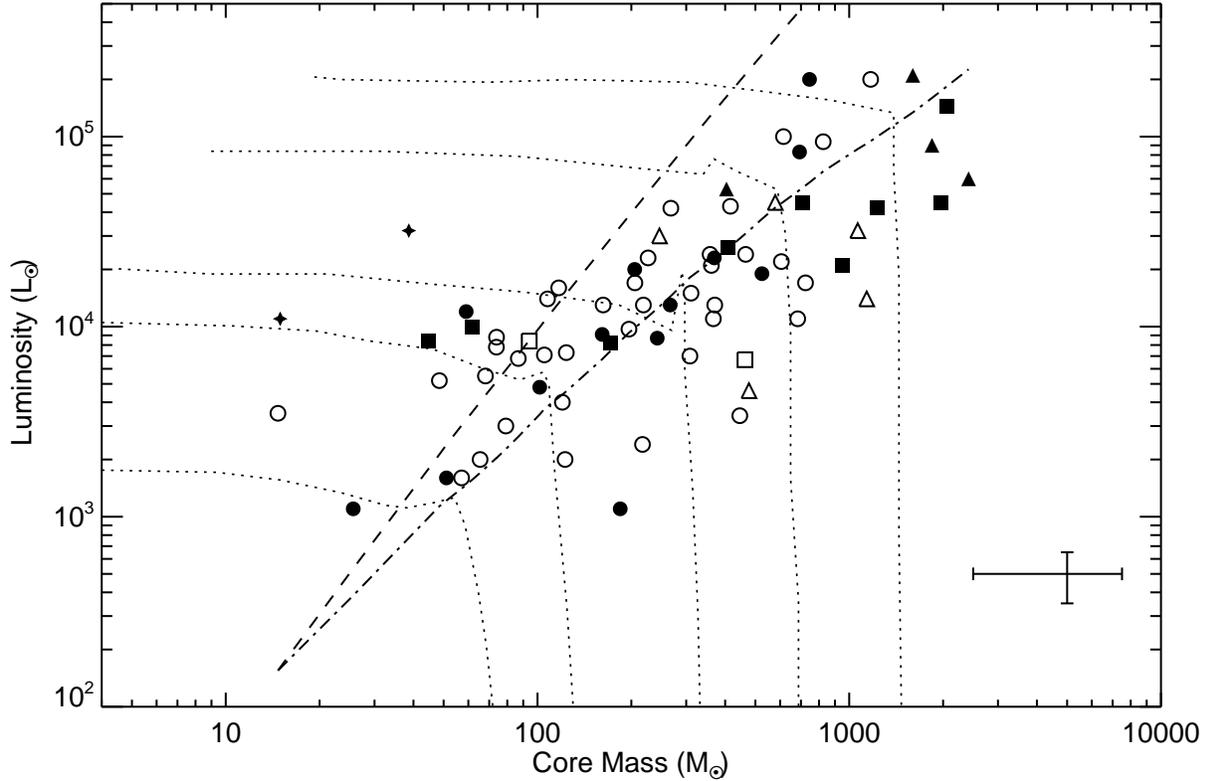}
\caption{Mass-luminosity plot for all MYSOs and H{\sc ii} regions, with mass flags 0 and 1 as circles, flag 2 sources as squares and flag 4 sources as triangles. Open and filled symbols separate MYSOs and H{\sc ii} regions. The two star symbols are the flag 7 H{\sc ii} regions G077.9637$-$00.0075 and G080.9383$-$00.1268 that appear to have dispersed their core material (assuming the cores were morphologically similar to the other cores in the sample at an earlier evolutionary stage). The dot-dashed line is the ZAMS luminosity expected from the most massive star in the core, when the stellar masses are distributed according to the Salpeter IMF in the case of a 50 percent SFE (the mass in stars equals the core mass). For the ZAMS line to fit the data, higher SFEs are required (moving it upwards). The dashed line is the luminosity of a cluster of ZAMS stars distributed with a Salpeter IMF with a 50 percent SFE. A reduced SFE is required if one were to use the luminosity from the whole cluster as it is already overestimates the luminosity of the most massive cores (see Section \ref{sfe}). The dotted lines represent the pre-main sequence tracks extracted from  models by \citet{Molinari2008} and can be used as a tool to trace evolution. Illustrative error bars of 50 percent in mass and 30 percent in luminosity are in the right-hand bottom corner.}
\label{fig:fig14} 
\end{center}
\end{figure*}

Relationships between mass and luminosity can provide insights into
star-formation efficiency and evolution in the cores.  In our case the gas
masses and source luminosities are independently established, whereas other works
 often obtain them via the same means, e.g. points in the mm/sub-mm SED.  The
downside to this is that many of the luminosities we have derived are at high spatial
resolution, generally of the dominant source present within the original MSX
beam.  Therefore the C$^{18}$O maps for such sources have effective radii larger than the 
``typical'' beamsize for the SED from which we derive the luminosity.  However, 
it is important to confirm that relationships found
between continuum mass and luminosities are not due to effectively comparing
the same data with itself \citep[e.g.][]{Molinari2008}. Figure \ref{fig:fig14}
plots RMS source luminosity against core gas mass for both MYSOs and H{\sc ii}
regions (open and filled symbols, respectively).  A linear relationship is
observed for both MYSOs and H{\sc ii} regions when plotting sources with mass
flags 0, 1 and 2 (and when including flag 4 sources).  The fitted
mass-luminosity relationship for MYSOs only is $L \propto M^{1.08\pm0.50}$ and
for H{\sc ii} regions only it is $L \propto M^{1.09\pm0.46}$ (using a bisector
fit that best represents the scatter in the data for flag 0, 1, 2 and 4
sources, as the latter follow the same trend). \citet{Urquhart2014b} find this
slope continues for $\sim$2 orders of magnitude in mass moving to much larger
clump scales (Radius $>$ 1\,pc).

\subsubsection{Probing core evolution}
\label{pcevo}

Evolutionary pre-main sequence tracks from \citet{Molinari2008}, based on the
model of \citet{Mckee2003}, are also shown in Figure \ref{fig:fig14}.  The basic
physics that underpins these tracks is relatively simple.  Protostars
continuously increase in luminosity, both through their strong
accretion phases and during Kelvin-Helmholtz (KH) contraction, until they reach
their eventual zero age main sequence (ZAMS) configuration (see, e.g.
\citealp{Hosokawa2009,Hosokawa2010,Zhang2014}).  Once on the main sequence, massive and
intermediate mass stars fairly quickly disperse their natal molecular material
through the action of their ionising radiation and winds.

We have also plotted the location of the stellar ZAMS in  Figure \ref{fig:fig14}. 
We take stellar luminosities from
\citet[][Figure 5.11]{Salaris2006} for masses ranging from 0.5\,M$_{\odot}$ to
6\,M$_{\odot}$ and from \citet{Davies2011} for masses $>$6\,M$_{\odot}$. We
assume that the total mass of stars $M_{\rm stars}$ = $M_{\rm core}$ and the
protostellar masses are distributed according to the Salpeter power-law IMF
(using only stellar masses from 0.5 to 150 M$_{\odot}$).  This assumption is
equivalent to a star-formation efficiency (SFE) of 50 percent, following
\citet{Lada2003} [SFE = $M_{\rm stars}$/($M_{\rm stars}$ $+$ $M_{\rm gas}$)], if no gas has
been lost due to winds or outflows.  This ZAMS line is shown as the dot-dashed
line on Figure \ref{fig:fig14}, and the evolutionary tracks as dotted lines, for initial
core masses of 80, 140, 350, 700 and 2000\,M$_{\odot}$.  The turnover in the
\citet{Molinari2008} tracks towards envelope dispersal occurs on this ZAMS line
as expected.  A change in the SFE is equivalent to shifting this ZAMS line
right or left (the luminosity does not change, but the residual core mass is
either greater for a lower SFE or less for a greater SFE).  This is discussed further
in the next section.  We have not attempted to model a cluster of
stars in detail, given that high mass stars evolve more quickly than lower mass
stars, and hence reach their ZAMS luminosity more quickly.  At least for
clusters of protostars this should mean that the higher mass members dominate
the luminosity (the same is not true for H{\sc ii} regions - \citealp{Lumsden2003}).

The relatively small scatter in this diagram suggests that most of the objects
we have observed are of similar evolutionary stage, as noted by \citet{Urquhart2014b}.
It is worth noting that
the H{\sc ii} regions overlap with the MYSOs, which suggests that our `compact' size
criterion ensures that for the most part we have selected relatively young H{\sc ii}
regions.  \citet{Molinari2008} note that the model PMS tracks are variable and
dependent upon accretion and core-dispersal rates, for example, such that they
can shift both vertically and horizontally which may partially explain this spread, as
could the observational errors in both mass and luminosity. Furthermore, these tracks assume
that the core does not gain mass from its surroundings by large-scale infall during star formation,
and that one core leads to one star with a fixed star formation efficiency. If, for example,
a core gains mass during the formation process, the tracks would also slant to the right as they increase
in luminosity as they gain \emph{extra} mass. An alternative
explanation however is given by the timescales involved for each phase of
evolution as a function of final mass.  The most massive stars reach the ZAMS
(and hence power H{\sc ii} regions) much more rapidly than it takes a less
massive star \citep[e.g][]{Urquhart2014b}.  Therefore the most massive stars reach the ZAMS and form H{\sc
  ii} regions whilst still very heavily embedded.  Notably, the H{\sc ii}
regions of lower mass appear more evolved on average in this figure than those
of higher mass, which suggests that this is what we are seeing. Although it is likely all
scenarios contribute to some extent. The filled
star symbols in Figure \ref{fig:fig14} show the two H{\sc ii} regions flagged
as 7 that are thought to be the most evolved and have dispersed their
cores. Their position to the far left hand side of the mass-luminosity diagram
supports this interpretation.

Both \citet{Mottram2011b} and \citet{Davies2011} find the lifetimes of MYSO and
H{\sc ii}-region phases to be roughly comparable (of the order
10$^{5}$\,yr). In particular, Figure 7 of \citet{Davies2011} illustrates how
the source classifications for stars with final masses $>$8\,M$_{\odot}$ change
(from MYSO to H{\sc ii} region) over a very narrow period in time where core
masses and luminosities are still coincident. As this paper is a pre-cursor to
a detailed investigation of molecular bipolar outflows, a narrow distribution
of evolutionary stage makes this sample ideal for identifying trends in outflow
parameters due to source properties, free from any effects of a spread in
source ages.

\begin{figure}
\begin{center}
\includegraphics[width=0.45\textwidth]{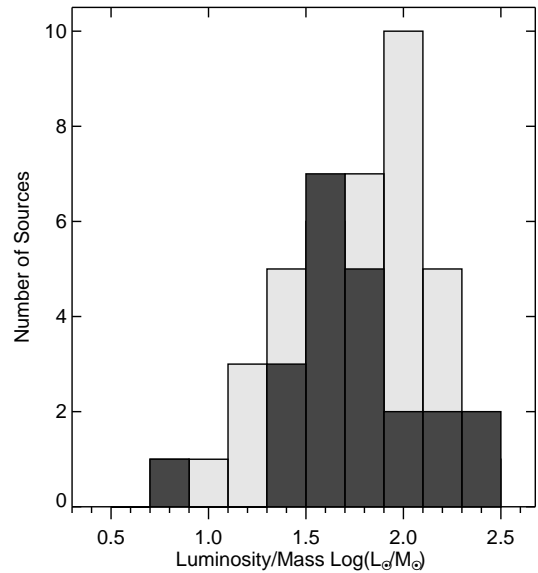} 
\caption{Histogram of the Luminosity/Mass ratio for MYSOs and H{\sc ii} regions, respectively, in light and dark grey bars. Both a KS and Mann-Whitney U test cannot distinguish MYSOs and H{\sc ii} regions as different distributions.}
\label{fig:fig15} 
\end{center}
\end{figure}

\citet{Elia2010} and \citet{Beltran2013} use the $L/M$ ratio as an evolutionary
tracer, where more evolved sources are more luminous and have dispersed more
material from their cores/envelopes (increasing $L/M$) in comparison with lower
luminosity, deeply embedded sources (the `evolved' flag 7 sources have $L/M$
$\gtrsim$ 10$^{3}$). \citet{Beltran2013} indicate a clear dichotomy between
24\,$\mu$m-bright and dark sources in their survey of the G29.96-0.02 cloud. In
a similar vein, the distribution of $L/M$ in the H{\sc ii} regions are
examined, looking for differences with MYSOs. Of course one caveat is that we
assume that to produce a source of a certain luminosity we always
start at the same mass (i.e. star formation efficiency does not vary). It is
not clear that this is the case from core to core, and therefore can introduce some scatter
in the distribution of sources (as previously discussed).
Figure \ref{fig:fig15} shows the
$L/M$ histogram for MYSOs and H{\sc ii} regions, and shows no clear difference
between the two subsamples. Both a Kolmogorov-Smirnov (KS) test and a
Mann-Whitney U (MWU) test indicate that MYSOs and H{\sc ii} regions are drawn
from the same distribution (as also found by \citealt{Urquhart2014b}).
The reported D value ($\sim$0.17) is not above that
required ($>$0.32) for the sources to be detectably drawn from different
distributions in the KS test. The reported MWU test probability ($\sim$0.46) is
well above the 0.05 significance level and so the hypothesis that the two
samples are drawn from the same distribution is not rejected.

\subsubsection{Star formation efficiency from cores to stars}
\label{sfe}

The masses plotted in Figure \ref{fig:fig14} are those of the clumps containing
what is probably a cluster of protostars within which the massive protostar
detected by the RMS survey is forming. However, the luminosities could either be dominated
by the most massive star, or be the total of all the protostars in the cluster.
The dot-dashed line in Figure \ref{fig:fig14} represents the luminosity of
a single star, the most massive in the cluster.  In principle, this line should set an upper
envelope to the observed data for single stars, and allow us to 
derive an estimate of the star formation efficiency (SFE).  As noted above,  although clusters of
protostars are probably dominated by their most massive member, the cluster line
still lies somewhat higher in luminosity than
the single star line (dashed line).  
The shape of the single star line matches the data reasonably well, suggesting
that this simple model is reasonable, i.e. the most massive star dominates. The cluster
  line however, is steeper. This is likely due to the fact that we plot ZAMS luminosities, which
  is a reasonable assumption for the more massive protostars, but probably underestimates the
  luminosity of accreting low mass protostars in theses clusters.

The overall SFE should be the value that sets all of our observations below the actual
model lines. If each of our objects genuinely only contained a single luminous
source, or else they are on the track beyond the ZAMS and into the core dispersal region, they will
fall below the dot-dashed line. This requires a shift to lower core masses (higher efficiencies) by close to a
factor of two. Such a large shift, and high SFE ($>$60\,percent), appears implausible.  Instead
the alternative that we are seeing clusters (dashed line), still in agreement with our
observations, which enhances the total luminosity at larger core masses by about a factor of two,
seems plausible.  In this case the SFE would be in the 40-50\,percent range.

\section{Summary}
\label{conc}

 The core parameters of 94 sources (of 99) MYSOs and H{\sc ii} regions selected
 as outflow candidates representative of the pre-2008 RMS survey, have been
 established. The sample is reduced to 89 when enforcing the distance limit of
 6\,kpc and is still representative of massive protostars across the MYSO to
 H{\sc ii} region transition in the up-to-date RMS survey that now meet the
 original selection criteria.

The majority of the cores exhibit a single, Gaussian C$^{18}$O line profile.
Larson-like relationships (FWHM linewidth relationships with luminosity, radius
and mass) are found for all the cores, though the scaling for these is not
continuous with other studies of larger-scale molecular clouds.  The most
fundamental relationship is between mass and core radius, which gives rise to
a surface density independent of radius.  We note, however, that the scatter in the
surface density in this relation
 is well explained by the correlation with gas opacity and
agrees with models in which Larson-style relations arise due to the observed
surface or column density limits.  We find two possible fundamental planes in
this work, representing core evolution (mass-luminosity-radius) and likely
virial contributions (mass-radius-FWHM).  All cores appear to be virialised
$M_{\rm core}$ $\sim$ $M_{\rm vir}$. Core parameters on the observed scales are
interpreted as being set prior to the onset of star formation and are not
subsequently effected by the feedback from the massive protostars.

A correlation is found between the dust-continuum masses and the core (gas)
masses established from the C$^{18}$O emission. At the resolution of the
observations, both the dust-continuum emission and C$^{18}$O emission trace the
same material and structures in both MYSO and H{\sc ii} regions.  The small
differences found are all consistent with dust being a better tracer of diffuse
low density gas on larger scales, whereas the C$^{18}$O traces the dense hearts
of the molecular clumps.

The sources are consistent with most of the luminosity arising in a single
massive protostar (or central ZAMS stars for the H{\sc ii} regions).  The tight
banding of all sources in the M-L plot, accounting for scatter, indicates a
similar evolutionary stage for both the MYSOs and H{\sc ii} regions
investigated.  This means they are ideal candidates to investigate
relationships between outflow and source properties between the source types to
be detailed in an upcoming paper.  

Further development of the RMS survey database is required with higher spatial
resolution at millimetre/sub-millimetre wavelengths to understand how the core
masses are specifically related to the immediate regions surrounding the most
massive protostars on arcsecond scales. This will allow us to investigate
whether the mass distributions are different for MYSOs and H{\sc ii} regions,
if CO depletion is present in these cores and on what spatial scale, whether
cores have further velocity sub-structure influenced by outflows, and how
Larson-type relationships hold below 0.1\,pc scales.

\section*{Acknowledgments}
Support for this work was in part provided by the Science and Technology Facilities Council (STFC) grant. The authors would like to 
thank the reviewer for their comments and suggestions that helped improve the clarity of the paper. Work undertaken in this paper made significant use of the \textsc{starlink} software package (http://starlink.eao.hawaii.edu/starlink). This paper made use of information from the Red MSX Source survey database at http://rms.leeds.ac.uk/cgi-bin/public/RMS\_DATABASE.cgi which was constructed with support from the Science and Technology Facilities Council of the UK. The James Clerk Maxwell Telescope has historically been operated by the Joint Astronomy Centre on behalf of the Science and Technology Facilities Council of the United Kingdom, the National Research Council of Canada and the Netherlands Organisation for Scientific Research.

\bibliographystyle{mn2e}

\appendix
\onecolumn
\section{Temperature, optical depth, column density and mass calculations}
\label{AppendixA}
In this appendix the column density and mass equations are derived following from the result of \citet{Garden1991}, except for the CO(3$-$2) transition. The total column density of a linear, rigid rotor molecule under conditions of local thermodynamic equilibrium (LTE), with the populations of all levels characterised by a single excitation temperature, $T_{\rm ex}$, is obtained from the integral of the optical depth over the line profile:

\begin{eqnarray}
\label{eqn_ntot}
N_{\rm tot} = \frac{3k}{8 \pi^{3}B \mu^{2}} \, \frac{{\rm exp}[hBJ(J+1)/kT_{\rm ex}]}{(J+1)} \, \frac{T_{\rm ex} + hB/3k}{[1-{\rm exp}(-h\nu /kT_{\rm ex})]} \int \tau_{\rm \upsilon}\, d\upsilon\; 
\end{eqnarray}

\noindent where $B$ is the rotational constant, $\mu$ is the permanent dipole moment of the molecule and $J$ is the rotational quantum number of the lower state, in this case $J=2$ for the CO(3$-$2) transition. $k$ and $h$ are the Boltzmann and Planck constants respectively.

Both the excitation temperature $T_{ex}$ and optical depth $\tau_{18}$ are solved for. Assuming both the source function and initial intensity are blackbodies at the respective temperatures $T_{\rm ex}$ and $T_{\rm cmb}$ = 2.73\,K, then

\begin{eqnarray}
\label{eqn1_core}
T_{\rm mb} = \frac{h\nu}{k} \left[ \frac{1}{{\rm exp}(h\nu/kT_{\rm ex})-1} - \frac{1}{{\rm exp}(h\nu/kT_{\rm cmb})-1} \right] \times \, \left[ 1 - {\rm exp} (-\tau)\right]\;,
\end{eqnarray}

\noindent for any line transition. From Equation \ref{eqn1_core} the excitation temperature, $T_{\rm ex}$, can be derived, provided the optical depth of the line is known. The $^{13}$CO observed simultaneously with C$^{18}$O should be optically thick at the locations of peak emission in our cores, such that as $\tau_{13} \rightarrow \infty$, $[ 1 - {\rm exp} (-\tau)] \rightarrow $ 1. In the case that C$^{18}$O is optically thin, we first calculate the optical depth of  $\tau_{13}$ directly from the ratio of the observed $^{13}$CO and C$^{18}$O antenna temperatures, in order to confirm $^{13}$CO is thick:

\begin{eqnarray}
\label{eqn2_core}
\frac{T_{\rm mb,13}}{T_{\rm mb,18}} \simeq \frac{1 - e^{-\tau_{13}}}{1 - e^{-\tau_{18}}} = \frac{1 - e^{-\tau_{13}}}{1 - e^{(-\tau_{13}/{\rm R})}}
\end{eqnarray}

\noindent where $R$ is abundance ratio of [$^{13}$CO]/[C$^{18}$O] derived from,

\begin{eqnarray}
\label{eqn3_core}
\frac{[^{16}{\rm O}]/[^{18}{\rm O}]}{[^{12}{\rm C}]/[^{13}{\rm C}]} = \frac{58.8 \times {D_{\rm gc}}+37.1}{7.5 \times {D_{\rm gc}}+7.6}
\end{eqnarray}

\noindent following \citet{Wilson1994} and $D_{\rm gc}$ is Galactocentric distance in kpc. 

The excitation temperature of $^{13}$CO is now calculated.  Rearranging Equation \ref{eqn1_core} and substituting in the numerically established optical depth, $\tau_{13}$,

\begin{eqnarray}
\label{eqn4_core}
T_{\rm ex} = \frac{15.86}{{\rm ln}\left[1 + 15.86 / \left( \{T_{\rm mb,13}/(1-{\rm exp} (-\tau_{13}))\} + 0.047 \right)\right]},
\end{eqnarray}

\noindent where ${h\nu}(^{13}{\rm CO})/{k}$ = 15.86\,K, with $\nu(^{13}{\rm CO})$ = 330.58\,GHz and $T_{\rm mb,13}$ is the main-beam brightness temperature of the $^{13}$CO emission. 

The C$^{18}$O excitation temperature $T_{\rm ex}$ is assumed to be the same as that of the $^{13}$CO line, and thus the optical depth, $\tau_{18}$ can now be calculated for the C$^{18}$O emission from: 

\begin{eqnarray}
\label{eqn5_core}
\tau_{18} = -{\rm ln} \, \bigg[  1 - \frac{T_{\rm mb,18}}{15.80/[{\rm exp}(15.80/T_{\rm ex})-1] - 0.045} \bigg]
\end{eqnarray}

Here the approximation for $ \int \tau_{\rm \upsilon}\, d\upsilon\ $ follows \citet{Buckle2010} for the case where $\tau \, \ne \,$0:

\begin{eqnarray}
\label{eqtr}
\int \tau_{\rm \upsilon}\, d\upsilon\, = \left[ \frac{h\nu}{k} \left( \frac{1}{{\rm exp}(h\nu/kT_{\rm ex})-1} - \frac{1}{{\rm exp}(h\nu/kT_{\rm cmb})-1} \right) \right]^{-1} \, \frac{\tau}{[ 1 - {\rm exp} (-\tau)]} \, \int T_{\rm mb}\, d\upsilon\,
\end{eqnarray}

The brightness temperature, $T_{\rm mb}$, is the antenna temperature of the telescope divided by the beam efficiency, T$^{*}_{\rm A}$/$\eta_{\rm mb}$, and corresponds to the Rayleigh-Jeans brightness of a source minus the brightness of the cosmic microwave background with temperature, $T_{\rm cmb}$ = 2.73\,K, over the beam. Combining equations  \ref{eqn_ntot} and \ref{eqtr}, in the limit where $T_{\rm ex}$ $\gg$ $T_{\rm cmb}$ results in the column density:

\begin{eqnarray}
\label{eqn_nave}
N = \frac{3k}{8 \pi^{3}B \mu^{2}} \, \frac{{\rm exp}[hBJ(J+1)/kT_{\rm ex}]}{(J+1)} \, \frac{1}{(h\nu/k)} \, \frac{T_{\rm ex} + hB/3k}{[{\rm exp}(-h\nu /kT_{\rm ex})]} \, \int T_{\rm mb}\, \frac{1}{[ 1 - {\rm exp} (-\tau)]}\, d\upsilon\
\end{eqnarray}

\noindent where the permanent dipole moment for C$^{18}$O is 0.1101 Debye \citep{Chackerian1983}. Conforming to cgs units typically used in such analysis $B$=58.14\,GHz, $k$=1.381$\times$10$^{-16}$\,erg\,K$^{-1}$, h=6.626$\times$10$^{-27}$\,erg\,s, $\nu$(C$^{18}$O) = 329.33055\,GHz, velocity $\upsilon$ is in k\,ms$^{-1}$, $\mu$(C$^{18}$O) = 0.1101 $\times$10$^{-18}$\,StatC\,cm (where 1\,statC = 1\,g$^{1/2}$\,cm$^{3/2}$\,s$^{−1}$ = 1\,erg$^{1/2}$\,cm$^{1/2}$), $\tau$ becomes $\tau_{18}$, the calculated optical depth of the C$^{18}$O line calcualted in Equation \ref{eqn5_core} and $T_{\rm ex}$ is the calculated excitation temperature from Equation \ref{eqn4_core}. The column density for the C$^{18}$O (3$-$2) transition is therefore:

\begin{eqnarray}
\label{eqn6_corea}
N({\rm C^{18}O}) = 5.0\, {\times}\, 10^{12} \, \frac{{\rm exp}(16.74/T_{\rm ex}) \, (T_{\rm ex} + 0.93)}{{\rm exp}(-15.80/T_{\rm ex})} \, \int T_{\rm mb} \, \frac{\tau_{18}}{[1 - {\rm exp} (-\tau_{18})]}\, d\upsilon\;,{\rm cm}^{-2}
\end{eqnarray}

\noindent The mass of a source can then be calculated from the column density via:

\begin{eqnarray}
\label{eqn_mass}
M_{\rm gas} = N({\rm CO}) \bigg[ \frac{{\rm H_2}}{\rm C^{18}O} \bigg] \mu_{g}\,m(_{\rm H_2})\Omega\,D^2
\end{eqnarray}

\noindent where $\mu_{g}$ = 1.36 is the total gas mass relative to H$_2$, the abundance ratio $[\rm H_2/C^{18}O]$ is a combination of ${\rm H_2}/{\rm ^{12}CO}$ = 10$^{4}$ and ${\rm ^{16}O}/{\rm ^{18}O}$ = 58.8 $\times$ $D_{\rm GC}$(kpc) $+$ 37.1 \citep{Wilson1994}, where $D_{\rm GC}$ is the Galactocentric distance and $D$ is the distance of the source to the Sun, both in kpc. $\Omega$ is the solid angle corresponding to the emission in one pixel of the maps used in this work. Thus including the conversion factors the core gas mass in solar masses (M$_{\odot}$) is calculated for every pixel of the map using Equation \ref{eqn_nave18}. The total core mass is the summation within the defined aperture encompassing the core, as described in Section \ref{meth_cores}. 

\begin{eqnarray}
\label{eqn_nave18}
M_{\rm gas}\,({\rm M}_{\odot}) = 2.5\, {\times}\, 10^{-12} \, \theta^2 (\arcsec)\, D^2 ({\rm kpc})\, \bigg[ \frac{\rm H_2}{\rm C^{18}O} \bigg] \, \frac{{\rm exp}(16.74/T_{\rm ex}) \, (T_{\rm ex} + 0.93)}{{\rm exp}(-15.80/T_{\rm ex})}  \, \int T_{\rm mb} \, \frac{\tau_{18}}{[1 - {\rm exp} (-\tau_{18})]}\, d\upsilon
\end{eqnarray}

\begin{figure}
\begin{center}
\includegraphics[width=0.40\textwidth]{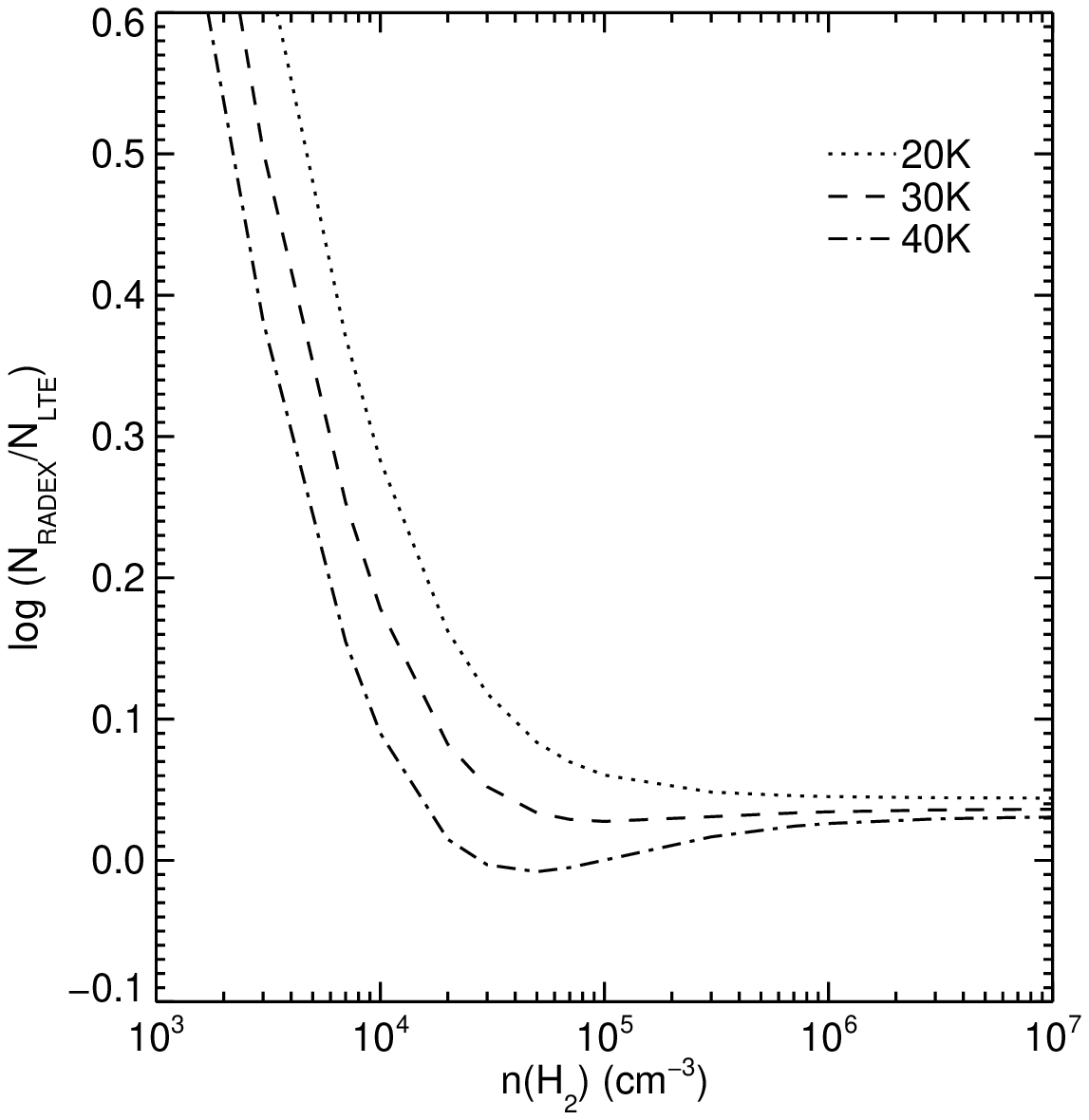}
\caption{Correction factor between calculated LTE calculated column density and that input into {\sc radex}. Although at 20\,K the correction factor is $\sim$2 at 10$^4$\,cm$^{-3}$, at the average volume density of the cores ($\sim$7$\times$10$^4$\,cm$^{-3}$) the correction factor is $\sim$ unity even at 20\,K.}
\label{fig:lte_lvg} 
\end{center}
\end{figure}

\noindent The LTE approximation is only valid if the volume density is sufficient enough (n$\gtrsim$n$_{\rm crit}$ = $\sim$3.2$\times$10$^4$\,cm$^{-3}$) \footnote{see, http://home.strw.leidenuniv.nl/$\sim$moldata/} such that the excitation temperature can be assumed to be equal to the kinetic temperature. If this is not the case (T$_{\rm ex}\,<$\,T$_{\rm kin}$) the LTE calculation will underestimate the column density and therefore mass. These parameters are then subject to a correction factor dependent on excitation temperature and volume density. Figure \ref{fig:lte_lvg} shows the correction factor as a function of volume density for kinetic temperatures of 20, 30 and 40\,K (i.e. the input excitation temperatures in LTE calculations). The factor is calculated by comparison of the input column density into {\sc radex} \citep{Vandertak2007} and that calculated via LTE calculation here. A representative H$_2$ column density of 2$\times$10$^{22}$\,cm$^{-2}$ and a linewidth of the average of all cores are used as inputs. At 20\,K for a 10$^4$\,cm$^{-3}$ volume density (just below critical), the correction factor required to obtain the correct column density from the presented LTE assumption is $\sim$2. Above the critical density the correction factor approaches unity. The average volume density of the cores however is $\sim$7$\times$10$^4$\,cm$^{-3}$ calculated using the LTE masses and deconvolved radii. Thus the LTE assumption is valid as the correction factor at this volume density is $\sim$ unity. There is no evidence for sub-thermal excitation of the CO $J$=3$-$2 transition.

\end{document}